\documentclass[aps, reprint, prb, twocolumn, amsmath, amssymb, groupedaddress, 10pt,
citeautoscript, longbibliography]{revtex4-1}

\usepackage{epsfig}
\usepackage{subfigure}
\usepackage{graphicx}
\usepackage{amsfonts}
\usepackage[figuresright]{rotating}
\usepackage{amssymb}
\usepackage{amsmath}
\usepackage{psfrag}
\usepackage{esint} 
\usepackage{hyperref}
\hypersetup{colorlinks,allcolors=blue}
\usepackage{multirow}
\usepackage{siunitx}
\usepackage{xcolor}
\usepackage{lipsum}
\usepackage{bbm}
\usepackage{soul}

\global\hyphenpenalty=100000

\begin{document}

\title{Spatially Inhomogeneous Triplet Pairing Order and Josephson Diode Effect Induced by Frustrated Spin Textures}

\author{Grayson R. Frazier}
\affiliation{Department of Physics and Astronomy, Johns Hopkins University, Baltimore, Maryland 21218, USA}
\author{Yi Li}
\affiliation{Department of Physics and Astronomy, Johns Hopkins University, Baltimore, Maryland 21218, USA}

\date{May 5, 2026}

\begin{abstract}
    We show that frustrated spin textures can generate anisotropic Josephson couplings between $d$ vectors that can stabilize spatially varying pairing orders in spin triplet superconductors.
    These couplings depend on the relative orientation of $d$ vectors, analogous to Dzyaloshinskii-Moriya and $\Gamma$-type interactions in magnetism, leading to an effective ``pliability'' of the pairing order that competes with superfluid stiffness.
    Such couplings cannot originate from spin-orbit coupling;
    rather, they can arise, for example, when itinerant electrons are coupled to a local exchange field composed of frustrated spin moments.
    Using a $T$-matrix expansion, we show that coupling to a local exchange field leads to an effective tunneling of itinerant electrons that is dependent on the underlying spin configurations at the barrier between superconducting grains.
    Furthermore, Josephson tunneling through frustrated spin textures can produce a Josephson diode effect.
    The diode effect originates either from nonvanishing spin chirality in the barrier, or from antisymmetric Josephson coupling between noncollinear $d$ vectors, both of which break inversion and time-reversal symmetries.
\end{abstract}

\maketitle

\section{Introduction}

The interplay between frustrated magnetism and unconventional superconductivity is an evolving
landscape in investigating strongly correlated systems.
Frustrated magnetic systems host noncollinear or noncoplanar spin textures which can lead to, for example, skyrmions, spin-momentum locking, and anomalous Hall effects~\cite{Ohgushi2000, Taguchi2001, Bruno2004, Metalidis2006, Xiao2010, Nagaosa2010, Nagaosa2012, Hellman2017}.
Spin triplet superconductivity is a class of unconventional superconductivity where Cooper pairs 
are characterized by an
internal degree of freedom encoded in the $d$ vector pairing order parameter~\cite{Leggett1975, Sigrist1991, Mackenzie2003, Cornfeld2021}.
In contrast to conventional superconductors, this degree of freedom allows direct interplay
and can open new avenues to studying the coexistence of frustrated spin textures and unconventional superconductivity~\cite{Linder2015, Eschrig2015, Frazier2025, Zhang2026}.

Recently, several systems hosting frustrated spin textures have been shown to exhibit unconventional superconductivity.
For example, spin triplet superconductivity has been shown to arise in proximitized kagome Weyl semimetal Mn$_3$Ge~\cite{Kiyohara2016, Liu2017, Wuttke2019, Dasgupta2020}, a system hosting $120^\circ$ ordered chiral antiferromagnetic spin configuration with an anomalous Hall effect in its normal state.
The helical spin texture resulting from coupling to local frustrated spins can lead to spin-valley locking, promoting spin triplet superconducting pairing correlations.
In proximity to superconducting Nb, the system shows long range coherent Josephson supercurrents~\cite{Jeon2021, Hou2025}, and under an out-of-plane magnetic field, the system produces hysteretic Josephson supercurrents, attributed to the finite spin chirality of the underlying noncoplanar spin texture~\cite{Jeon2023}.

Additionally, there has been proposed interplay between frustrated spin textures and unconventional superconductivity in transition-metal dichalcogenide $4H_b$-TaS$_2$, consisting of alternating layers of centrosymmetric spin liquid candidate layer $1T$-TaS$_2$ and noncentrosymmetric superconducting layer $1H$-TaS$_2$.
The material exhibits
chiral superconducting states, spontaneous vortices, and the ``magnetic memory" effect in addition to spin triplet pairing~\cite{Ribak2020, Persky2022, Silber2024}.
Recent work proposes that frustrated magnetic textures underlie the observed phenomena~\cite{Lin2024, Liu2024, Koenig2024, Crippa2024, Levitan2025}.
However, the role of frustrated spin textures and how they can affect unconventional pairing order remains an open avenue to explore.


It has recently been proposed that frustrated spin textures can give rise to anisotropic Josephson couplings favoring spatially inhomogeneous spin triplet pairing order~\cite{Frazier2025}.
In the present work, we expand upon this theoretical approach, considering a momentum-dependent $d$ vector and incorporating the effects of spin-orbit coupling and a frustrated local exchange field.
In Sec.~\ref{sec:josephson_coupling}, we propose the free energy for a spatially varying spin triplet pairing order and demonstrate how it microscopically originates from the Josephson coupling.
The anisotropic Josephson coupling terms cannot emerge from spin-orbit coupling but rather arise when itinerant electrons are coupled to frustrated local spin moments.
To illustrate, we analyze a three-sublattice system on geometrically frustrated lattices in Sec.~\ref{sec:models_and_tunneling}, in which $s$-$d$ exchange gives rise to an effective tunneling dependent on the underlying spin structure.
In Sec.~\ref{sec:pairing_correlations},
we analyze the spin triplet pairing correlations for an isolated superconducting grain arising from $s$-$d$ exchange or spin-orbit coupling.
Next, accounting for the effective tunneling in the presence of a frustrated local exchange field, we derive the effective Josephson couplings for the three-sublattice system in Sec.~\ref{sec:josephson_three_sublattice}.
The emergence of anisotropic Josephson coupling leads to a ``pliability'' of the pairing order that competes with the superfluid stiffness and can promote a spatially varying spin triplet pairing order, as discussed in Sec.~\ref{sec:competition_of_bulk_and_josephson_coupling}.
Lastly, in Sec.~\ref{sec:josephson_diode}, we demonstrate that when the underlying spin structure has finite spin chirality, or when there is antisymmetric Josephson coupling between noncollinear $d$ vectors, the system exhibits a Josephson diode effect.

\section{Free energy contribution from spatially varying pairing order}

\label{sec:josephson_coupling}

Generally, a system will adjust the superconducting pairing order to minimize the total free energy, which can be decomposed into two principal contributions,
\begin{equation}
    F = F_\mathrm{homogeneous} + F_\mathrm{variation}.
    \label{free_energy_general}
\end{equation}
The first term describes the bulk pairing order for a single isolated homogeneous superconducting grain and is associated with the pairing condensation energy, while the latter term describes the contribution from a spatially varying pairing order, which can either promote or penalize spatial inhomogeneity. 
In this work, we focus on the contribution to the free energy from a spatially varying pairing order for a spin triplet superconductor, deriving anisotropic Josephson couplings which can favor a spatially inhomogeneous spin triplet pairing order.
We consider sufficiently large superconducting grains, or an emergent network of Josephson junctions in a single-crystal~\cite{Yasui2020, Blom2025}, in which the Coulomb charging energy is negligible.

\begin{figure}
    \centering
    \includegraphics[width=\linewidth]{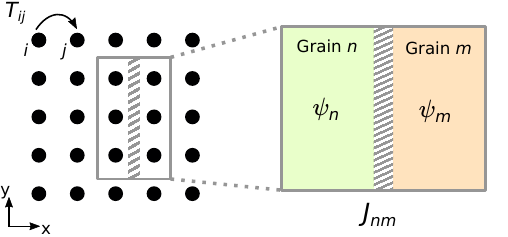}
    \caption{
    Discretized Josephson free energy describing the contribution from a spatially varying order.
    We consider a system consisting of multiple weakly-linked superconducting grains.
    For an $s$-wave superconductor, the pairing correlation defined locally for the $n\mathrm{th}$ superconducting grain is given by $\psi_n$.
    The discretized superfluid stiffness $J_{nm}$ describes the Josephson coupling between grains $n$ and $m$.
    }
    \label{fig:josephson_discrete}
\end{figure}

We first review the kinetic energy contribution to the free energy,
which, in the discrete limit, can be related to the Josephson coupling.
Consider a superconductor with a spatially varying pairing order.
For a conventional $s$-wave superconductor with complex scalar pairing order parameter $\psi(\mathbf{r})$, the energy cost from a spatially varying order in the absence of an external field is given by
$F_\mathrm{variation} = \int \mathrm{d}^dr \, \gamma |\boldsymbol{\nabla} \psi(\mathbf{r})|^2$, with $\gamma$ related to the superfluid stiffness.
In the discrete limit, the gradient term corresponds to the Josephson coupling between superconducting grains~\cite{Spivak1991, Sigrist1992, Sigrist1995, Kapitulnik2019, Zhang2025b},
\begin{math}
    F_\mathrm{variation} = \sum_{\langle n m \rangle}
    J_{nm} \psi_n \psi_m^* + \mathrm{c.c.}
\end{math}
Here, neighbouring superconducting grains $n$ and $m$ described by pairing order parameters $\psi_n$ and $\psi_m$ interact via Josephson coupling at their interface, with the amplitude of the Josephson coupling given by $J_{nm}$, as shown in Fig.~\ref{fig:josephson_discrete}.
Physically, this coarse-grained picture corresponds to weakly-coupled superconducting grains, where, within each superconducting grain, on the order of the coherence length of the Cooper pair, the local pairing order can be treated as uniform.
In the absence of external fields or disorder, the lattice version of the superfluid stiffness $J_{nm}$ is typically negative valued.
As such, the free energy is minimized when $\psi_n$ and $\psi_m$ have the same $\mathrm{U}(1)$ phase, leading to a phase-coherent and spatially homogeneous pairing order for sufficiently large grains.

\subsection{Free energy contribution for spin triplet pairing order}

\begin{figure}
    \centering
    \includegraphics[width= \linewidth]{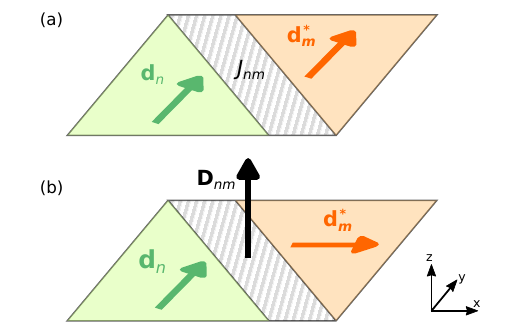}
    \caption{Josephson couplings between superconducting grains $n$ and $m$ described by $d$ vectors $\hat{\mathbf{d}}_n$ and $\hat{\mathbf{d}}_m$.
    \textbf{(a)}~Heisenberg-like coupling, corresponding to the discretized superfluid stiffness, which leads to collinear $d$ vector configurations.
    \textbf{(b)}~DM-like Josephson coupling, which can lead to noncollinear $d$ vector textures, leading to an effective pliability in the superconducting order.
    }
    \label{fig:Jnm_and_Dnm_schematic}
\end{figure}

We now consider spin triplet superconductors, in which, in addition to a $\mathrm{U}(1)$ phase, the orbital angular momentum and spin of the Cooper pair are also degrees of freedom that contribute to the free energy.
For superconducting grain $n$, the pairing order is described by its overall phase $\phi_n$ and $d$ vector ${\mathbf{d}}_n(\mathbf{k})$~\cite{Balian1963, Leggett1975, Volovik2009}.
In general, the $d$ vector can be momentum-dependent, with $\mathbf{k}$ describing the relative momentum of the Cooper pair.
The first order Josephson coupling between grains $n$ and $m$ includes all allowed quadratic couplings of the $d$ vectors and takes the following form,
\begin{align}
    &F_{nm}(\mathbf{k}, \mathbf{k}')
    = \nonumber
    e^{i(\phi_n - \phi_m)}
    \Big\{
    J_{nm}(\mathbf{k}, \mathbf{k}')
    {\mathbf{d}}_n(\mathbf{k})
    \cdot {\mathbf{d}}_m^*(\mathbf{k}')
    \\
    &\hspace{2em}+ \nonumber
    \mathbf{D}_{nm}(\mathbf{k}, \mathbf{k}')
    \cdot
    \Big(
    {\mathbf{d}}_{n}(\mathbf{k}) \times {\mathbf{d}}_m^*(\mathbf{k}')
    \Big)
    \\
    &\hspace{2em}+ 
    \sum_{a,b = 1,2,3}
    {d}_n^a(\mathbf{k}) \Gamma_{nm}^{ab}(\mathbf{k}, \mathbf{k}')
    {d}^{*b}_m(\mathbf{k}')
    \Big\}
    +
    \mathrm{c.c.}
    \label{free_energy_josephson_coupling_discrete}
\end{align}
Above, ${\mathbf{d}}_n(\mathbf{k})$ and $\phi_n$ are the $d$ vector and overall $\mathrm{U}(1)$ phase of the pairing order of the $n\mathrm{th}$ grain.
The three terms in the above free energy can be understood in analogy to 
superexchange~\cite{Dzyaloshinsky1958, Moriya1960, Moriya1960a, Hellman2017, Hill2021} in classical spin systems as follows.
The first term $J_{nm}$ corresponds to a Heisenberg-like symmetric coupling which favors collinear alignment of $d$ vectors, whereas the second term $\mathbf{D}_{nm}(\mathbf{k}, \mathbf{k}')$ is an antisymmetric Dzyaloshinskii-Moriya (DM)-like coupling that favors noncollinear textures, {as shown in Figs.~\ref{fig:Jnm_and_Dnm_schematic}(a) and (b) respectively.}
The third term $\Gamma_{nm}(\mathbf{k}, \mathbf{k}')$ is a symmetric traceless matrix corresponding to ``$\Gamma$-type'' exchange~\cite{Rau2014}.

Minimization of the free energy in Eq.~\eqref{free_energy_josephson_coupling_discrete}
with respect to the three types of Josephson couplings can lead to a spatially inhomogeneous $d$ vector texture.
The total contribution to the free energy $F_\mathrm{variation} = \sum_{\mathbf{k}, \mathbf{k}'} \sum_{\langle n m \rangle} F_{nm}(\mathbf{k}, \mathbf{k}')$
is given by
summing over the relative momenta $\mathbf{k}$ and $\mathbf{k}'$ of all neighbouring grains $n$ and $m$, respectively.
In the continuum limit, the free energy is given by $F_\mathrm{variation} = \int \mathrm{d}^d r 
f_\mathrm{variation} (\mathbf{r})$,
in which the free energy density contains the following gradient terms,
\begin{align}
    f_\mathrm{variation} (\mathbf{r})
    &= \nonumber
    \sum_i \gamma_i(\mathbf{r}) 
    \partial_i {\mathbf{d}}(\mathbf{r})
    \cdot
    \partial_i{\mathbf{d}}^*(\mathbf{r})
    \\
    &+ \nonumber
    \sum_i \mathbf{A}_i (\mathbf{r}) \cdot (
    \partial_i {\mathbf{d}}(\mathbf{r})
    \times
    {\mathbf{d}}^*(\mathbf{r})
    )
    \\
    &+
    \sum_{ij}
    \sum_{ab}
    K_{ij}^{ab}(\mathbf{r})
    \partial_i d^{a}(\mathbf{r})
    \partial_j {d}^{b*}(\mathbf{r})
    + \mathrm{c.c.}
    \label{Josephson_free_energy_continuum}
\end{align}
Above, we have absorbed the $\mathrm{U}(1)$ phase into the definition of the $d$ vector and integrated out the $k$-dependence.
%
The coupling amplitudes
$\gamma_i(\mathbf{r})$, $\mathbf{A}_i(\mathbf{r})$, and $K_{ij}^{ab}(\mathbf{r})$ are the continuum analogues of $J_{nm}$, $\mathbf{D}_{nm}$, and $\Gamma_{nm}$, respectively.
$J_{nm}(\mathbf{k}, \mathbf{k}')$ favors collinear $d$ vector textures, and when it is negative-valued, it can stabilize phase-coherent and spatially homogeneous $d$ vector textures.
However, the presence of finite $\mathbf{D}_{nm}(\mathbf{k}, \mathbf{k}')$ and $\Gamma_{nm}(\mathbf{k}, \mathbf{k}')$ can favor spatially frustrated $d$ vector textures, leading to, for example, the formation of vortices even in the absence of external fields~\cite{Frazier2025}.
This leads to an effective ``pliability'' of the pairing order that competes with the discrete superfluid stiffness $J_{nm}$ and homogeneous contribution $F_\mathrm{homogeneous}$, such that the system tends towards spatially nonuniform $d$ vector textures.

{
We note that the DM-like couplings here can be viewed as a particular realization of a Lifshitz invariant---terms in the Ginzburg-Landau free energy which are linear in spatial gradients, as seen in Eq.~\eqref{Josephson_free_energy_continuum}.
Such terms are closely related to the Lifshitz invariants discussed in, for example, Refs.~\citenum{Mineev1994} and \citenum{Mineev2008}, in which spin-orbit coupling can lead to antisymmetric gradient terms in the free energy for multicomponent superconductors and can stabilize helical superconducting textures.
In contrast, the mechanism in the present work does not rely on spin-orbit coupling nor a noncentrosymmetric crystal; rather, we demonstrate in the following sections that frustrated spin textures generate anisotropic Josephson couplings between $d$ vectors, leading to Lifshitz-type gradient terms which originate from DM-like Josephson couplings.
Furthermore, we discuss the microscopic origin of all possible symmetry-allowed interchannel couplings seen in Eq.~\eqref{free_energy_josephson_coupling_discrete}, including antisymmetric and symmetric traceless contributions.
The resulting spatially varying pairing order  can originate from magnetic frustration rather than spin-orbit effects, providing a distinct route towards engineering inhomogeneous superconducting textures.
}

\subsection{Microscopic derivation of Josephson couplings}
\label{subsec:josephson_microscopic}

The Josephson free energy in Eq.~\eqref{free_energy_josephson_coupling_discrete} can be derived from the Ambegaokar-Baratoff formalism of the first order Josephson coupling.
Consider two superconducting grains connected by a weak link.
Restricting to the orbital and spin degrees of freedom,
the effective single-particle tunneling across the barrier between grains $n$ and $m$ is given by~\cite{Bardeen1961, Cohen1962, Caroli1971, Caroli1971a}
\begin{equation}
    [H_{T}]_{n m} = 
    \sum_{\mathbf{k}, \mathbf{k}'; \alpha, \alpha'} \left(  c^\dagger_{n, \mathbf{k}, \alpha} [T_{n m}(\mathbf{k}, \mathbf{k}')]_{\alpha \alpha'} c_{m, \mathbf{k}', \alpha'}
    + \mathrm{h.c.} \right),
\end{equation}
in which $c_{n, \mathbf{k}, \alpha}$ is the annihilation operator for an electron in grain $n$ with momentum $\mathbf{k}$ and spin $\alpha = {\uparrow, \downarrow}$.
The effective tunneling matrix element between grains $n$ and $m$ generally takes the form
\begin{math}
    T_{n m}(\mathbf{k}, \mathbf{k}')
    = T_{n m; 0}(\mathbf{k}, \mathbf{k}')  \sigma^0 + \mathbf{T}_{n m}(\mathbf{k}, \mathbf{k}')  \cdot \boldsymbol{\sigma},
\end{math}
in which $\boldsymbol{\sigma} = (\sigma^x, \sigma^y, \sigma^z)^\mathrm{T}$ are the Pauli matrices.
Tunneling amplitudes $T_{n m; 0}(\mathbf{k}, \mathbf{k}')$ and 
$\mathbf{T}_{n m}(\mathbf{k}, \mathbf{k}') $ correspond to spin-independent and spin-dependent tunneling processes at the grain boundary, respectively.
The latter can arise from, for example, spin-orbit coupling or time-reversal breaking fields at the interface between superconducting grains.

The contribution to the Josephson free energy between neighbouring grains $n$ and $m$
is
\begin{equation}
    F_{nm}(\mathbf{k}, \mathbf{k}') = \frac{1}{2}( \mathcal{J}_{nm}(\mathbf{k}, \mathbf{k}')
    +
    \mathrm{c.c.}),
\end{equation}
in which the Josephson form factor $\mathcal{J}_{nm}$ is given by~\cite{Ambegaokar1963, Geshkenbein1986, Sigrist1991, Frazier2024}
\begin{align}
    \nonumber
    \mathcal{J}_{nm}(\mathbf{k}, \mathbf{k}')
    =
    - \frac{1}{\beta}
    & \sum_{i\omega_n}
    \mathrm{Tr}
    \Big[ \mathcal{F}_{n}(\mathbf{k}; i\omega_n) [T_{n m}(-\mathbf{k}, -\mathbf{k}'; i\omega_n)]^\mathrm{T}
    \\
    & \times
    [\mathcal{F}_{m}^\dagger(\mathbf{k}'; i\omega_n)]^\mathrm{T} T_{n m}(\mathbf{k}, \mathbf{k}'; i\omega_n)
    \Big].
    \label{josephson_form_factor}
\end{align}
Above, the trace is taken over internal degrees of freedom of the Cooper pair (\textit{e.g.} spin, sublattice, \textit{etc}.), and the summation is taken over fermionic Matsubara frequencies $i\omega_n = (2n + 1)\pi/\beta$ for integers $n$.
Here, 
\begin{math}
    [\mathcal{F}_{m} ]_{\alpha \beta} (\mathbf{k}; i\omega_n) = - \int \mathrm{d} \tau  e^{i\omega_n \tau} \langle \mathcal{T}_\tau c_{m, -\mathbf{k}, \beta} (\tau) c_{m, \mathbf{k}, \alpha}(0) \rangle
\end{math}
is the anomalous Green's function of grain $m$ that describes the pairing correlations.
The anomalous Green's function can be decomposed into its spin singlet and spin triplet parts as
\begin{equation}
    \mathcal{F}_{n}(\mathbf{k}; i\omega_n)
    =
    \Big( f_{n, 0} (\mathbf{k}; i\omega_n) + \mathbf{f}_n(\mathbf{k}; i\omega_n) \cdot \boldsymbol{\sigma} \Big)
    i\sigma^y,
\end{equation}
with $\mathbf{f}_n(\mathbf{k}; i\omega_n)$ playing the role of a $d$ vector describing spin triplet correlations.
Substituting into Eq.~\eqref{josephson_form_factor} yields three distinct contributions corresponding to the Josephson tunneling between spin singlet components, between spin triplet components, and between mixed spin singlet and spin triplet components, given in Appendix~\ref{appendix:Josephson_form_factor}.

Here, we focus on the Josephson form factor describing the couplings between spin triplet pairing orders of grains $n$ and $m$, given by
\begin{widetext}
\begin{equation}
    \begin{aligned}
        \mathcal{J}_{nm}^{\mathrm{trip}-\mathrm{trip}}(\mathbf{k}, \mathbf{k}')
        = 
        - \frac{1}{\beta}
        e^{i(\phi_n - \phi_m)}
        \sum_{i\omega_n}
        \Big(
        \mathcal{J}_{nm}^J(\mathbf{k}, \mathbf{k}')
        +
        \mathcal{J}_{nm}^\mathrm{DM}(\mathbf{k}, \mathbf{k}')
        +
        \mathcal{J}_{nm}^\Gamma(\mathbf{k}, \mathbf{k}')
        \Big).
    \end{aligned}
    \label{triplet_triplet_josephson_form_factor.condensed}
\end{equation}
The three contributions, corresponding to Heisenberg-like, DM-like, and $\Gamma$-type couplings of $d$ vectors, are respectively given by
\begin{equation}
    \begin{aligned}
        \mathcal{J}_{nm}^J(\mathbf{k}, \mathbf{k}') 
        &= 
        2 T_{n m; 0}(-\mathbf{k}, -\mathbf{k}')  T_{n m; 0}(\mathbf{k}, \mathbf{k}') \Big[ \mathbf{f}_n(\mathbf{k}') \cdot \mathbf{f}_m^*(\mathbf{k})\Big]
        + 
        2\Big[ \mathbf{T}_{n m}(-\mathbf{k}, -\mathbf{k}') \cdot \mathbf{T}_{n m}(\mathbf{k}, \mathbf{k}') \Big]
        \Big[ \mathbf{f}_n(\mathbf{k}') \cdot\mathbf{f}_m^*(\mathbf{k}) \Big],
        \\
        \mathcal{J}_{nm}^\mathrm{DM} (\mathbf{k}, \mathbf{k}')
        &=
        2i
        T_{n m; 0}(\mathbf{k}, \mathbf{k}') \mathbf{T}_{n m}(-\mathbf{k}, -\mathbf{k}') \cdot \Big[ \mathbf{f}_n(\mathbf{k}') \times \mathbf{f}_m^*(\mathbf{k}) \Big]
        +
        2i
        T_{n m; 0}(-\mathbf{k}, -\mathbf{k}')  \mathbf{T}_{n m}(\mathbf{k}, \mathbf{k}') \cdot \Big[ \mathbf{f}_n(\mathbf{k}') \times \mathbf{f}_m^*(\mathbf{k})  \Big],
        \\
        \mathcal{J}_{nm}^\Gamma (\mathbf{k}, \mathbf{k}')
        &= 
        -2
        \Big[\mathbf{f}_n(\mathbf{k}') \cdot \mathbf{T}_{n m}(-\mathbf{k}, -\mathbf{k}')\Big]
        \Big[\mathbf{f}_m^*(\mathbf{k}) \cdot \mathbf{T}_{n m}(\mathbf{k}, \mathbf{k}')\Big]
        -2
        \Big[\mathbf{f}_n(\mathbf{k}') \cdot \mathbf{T}_{n m}(\mathbf{k}, \mathbf{k}')\Big]
        \Big[ \mathbf{f}_m^*(\mathbf{k}) \cdot \mathbf{T}_{n m}(-\mathbf{k}, -\mathbf{k}')\Big].
    \end{aligned}
    \label{josephson_couplings_expanded}
\end{equation}
\end{widetext}
Above, the Heisenberg-like Josephson coupling $\mathcal{J}_{nm}^J$ can be realized for purely spin-independent tunneling; however, the DM-like and $\Gamma$-type interactions are respectively linear and quadratic in spin-dependent tunneling amplitude.
To realize the DM-like antisymmetric term $\mathcal{J}_{nm}^\mathrm{DM}$, it is necessary and sufficient for the effective tunneling to satisfy
\begin{equation}
    T_{nm;0}(\mathbf{k}, \mathbf{k}')
    \mathbf{T}_{nm}(-\mathbf{k}, -\mathbf{k}')
    \neq
    - T_{nm;0}(-\mathbf{k}, -\mathbf{k}')
    \mathbf{T}_{nm}(\mathbf{k}, \mathbf{k}').
    \label{dm_coupling_condition}
\end{equation}
Hence, it is sufficient for the spin-independent and spin-dependent tunneling $T_{nm;0}(\mathbf{k}, \mathbf{k}')$ and $\mathbf{T}_{nm}(\mathbf{k}, \mathbf{k}')$ to be even
under $(\mathbf{k}, \mathbf{k}') \rightarrow (-\mathbf{k}, -\mathbf{k}')$
for the DM-like Josephson coupling to be nonvanishing.
The antisymmetric Josephson coupling cannot arise, for example, through spin-orbit coupling alone, under which $\mathbf{T}_{nm}(\mathbf{k}, \mathbf{k}') = -\mathbf{T}_{nm}(-\mathbf{k}, -\mathbf{k}')$ for real-valued $\mathbf{T}_{nm}(\mathbf{k}, \mathbf{k}')$ due to time reversal symmetry.
For example, for Rashba-type spin-orbit coupling at the barrier, $\mathbf{T}_{nm}(\mathbf{k}, \mathbf{k}') = |T_{nm}| \delta_{\mathbf{k}, \mathbf{k}'} (\hat{\mathbf{n}} \times \hat{\mathbf{k}}) $, with $\hat{\mathbf{n}}$ the vector normal to the interface~\cite{Geshkenbein1986, Leggett2020, Frazier2024}, the DM-like Josephson coupling is vanishing.
Rather, the tunneling requires, for example, the presence of a local exchange field or impurities at the interface that break time reversal symmetry.

\subsubsection{System with decoupled spin and orbital degrees of freedom}

\label{subsec:josephson_coupling.decoupled_spin_orbital}

We first consider the case where spin and orbital degrees of freedom are decoupled, such as in a generalized $^3$He-A type pairing.
Suppose the pairing correlator takes the form
\begin{equation}
    \mathcal{F}_{m}^{(\mathrm{trip})}(\mathbf{k}; i\omega_n) = - \frac{\Delta_0}{\omega_n^2 + E_{m, \mathbf{k}}^2} g_m(\mathbf{k}) \hat{\mathbf{d}}_m \cdot \boldsymbol{\sigma} (i\sigma^y),
    \label{decoupled_spin_orbital_anomalous_GF}
\end{equation}
in which $\hat{\mathbf{d}}_m$ is momentum-independent, and $g_m(\mathbf{k})$ describes the orbital structure.
We consider tunneling of the form $T_{nm}(\mathbf{k}, \mathbf{k}') = T_{nm} h(\mathbf{k}, \mathbf{k}')$, with the dimensionless function $h(\mathbf{k}, \mathbf{k}')$ describing the momentum-dependence, with $h(\mathbf{k}, \mathbf{k}) = h(-\mathbf{k}, -\mathbf{k})$.
The Josephson free energy describing the coupling of grains $n$ and $m$ is given by
\begin{eqnarray}
        F_{n m}
        &=&
        e^{i(\phi_n - \phi_m)}
        \Big\{ J_{n m} \hat{\mathbf{d}}_{n} \cdot \hat{\mathbf{d}}_{m}^*
        +
        \mathbf{D}_{n m} \cdot \Big( \hat{\mathbf{d}}_{n} \times \hat{\mathbf{d}}_{m}^* \Big)
        \nonumber \\
        &&+
        \sum\nolimits_{a,b = 1,2,3}
        \hat{d}_{n}^{a} \Gamma_{n m; ab} \hat{d}^{*b}_{m}
        \Big\} 
        + \mathrm{c.c.},
\end{eqnarray}
with coupling coefficients
\begin{equation}
    \begin{aligned}
        {J}_{n m} &=
        w_{nm}(\beta)
        \left(
        T_{n m; 0}^2
        +
        \mathbf{T}_{n m} \cdot \mathbf{T}_{n m}
        \right),
        \\
        {\mathbf{D}}_{n m} &=
        2i 
        w_{nm}(\beta)
        T_{n m; 0} \mathbf{T}_{n m},
        \\
        {\Gamma}_{n m}^{ab} &= 
        -2 
        w_{nm}(\beta)
        T_{n m}^{a} T_{n m}^b.
        \label{josephson_couplings_decoupled_spin_orbital_expansion}
    \end{aligned}
\end{equation}
Here, the weighting factor is given by
\begin{align}
    &\nonumber
    w_{nm}(\beta)
    = 
    \Delta_0^2
    \sum_{\mathbf{k}, \mathbf{k}'}
    \bigg\{
    \frac{g_n(\mathbf{k})
    g_m^*(\mathbf{k}')
    h^2(\mathbf{k}, \mathbf{k}')
    }{2E_{n, \mathbf{k}} E_{m, \mathbf{k}'}}
    \times
    \\
    &
    \left(
    \frac{n_F(E_{n, \mathbf{k}}) + n_F(E_{m, \mathbf{k}'}) - 1}{E_{n, \mathbf{k}}+ E_{m, \mathbf{k}'}}
    -
    \frac{n_F(E_{n, \mathbf{k}}) - n_F(E_{m, \mathbf{k}'})}{E_{n, \mathbf{k}} - E_{m, \mathbf{k}'}}
    \right)
    \bigg\}
    ,
    \label{weighting_factor.sum_over_k}
\end{align}
with $n_F(E) = (1 + e^{\beta E})^{-1}$, and at
zero temperature, this reduces to
\begin{align}
    w_{nm}(\beta \rightarrow \infty)
    = 
    -\Delta_0^2
    \sum_{\mathbf{k}, \mathbf{k}'}
    \frac{g_n(\mathbf{k})
    g_m^*(\mathbf{k})
    h^2(\mathbf{k}, \mathbf{k}')}{2E_{n, \mathbf{k   }} E_{m, \mathbf{k}'} (E_{n, \mathbf{k}}+ E_{m, \mathbf{k}'})}.
    \label{w_factor_zero_temp}
\end{align}
Up to a factor from the overlap integral of $g_n(\mathbf{k})$ and $g_m(\mathbf{k})$,
the weighting factor scales approximately as
\begin{math}
    w_{nm} \sim 
    - \Delta_0/W^2,
\end{math}
in which $W$ is the bandwidth, as shown in Appendix~\ref{appendix:weighting_factor}.

\subsubsection{System with coupled spin and orbital degrees of freedom}

We next consider the case in which the spin and orbital degrees are coupled in the pairing order.
For simplicity, suppose
that the anomalous Green's function for the $m\mathrm{th}$ grain takes the form
\begin{equation}
    \mathcal{F}^{(\mathrm{trip})}_m(\mathbf{k}; i\omega_n)
    =
    - \frac{\Delta_0}{\omega_n^2 + E_{m, \mathbf{k}}^2}
    K_m(k)
    \hat{\mathbf{d}}_m(\hat{\mathbf{k}})\cdot \boldsymbol{\sigma} (i\sigma^y).
\end{equation}
Here, $\Delta_0$ is the magnitude of the superconducting gap, $E_{m, \mathbf{k}}$ is the Bogoliubov-de Gennes (BdG) quasiparticle energy, $K_m(k)$ is a scalar function of ${k=|\mathbf{k}|}$, and 
$\hat{\mathbf{d}}_m(\hat{\mathbf{k}})$ is a momentum-dependent $d$ vector satisfying ${\hat{\mathbf{d}}_m(\hat{\mathbf{k}}) = - \hat{\mathbf{d}}_m(-\hat{\mathbf{k}})}$.

For a two-dimensional system, 
the Heisenberg-like, DM-like, and $\Gamma$-type couplings contributing to the Josephson form factor can respectively be expressed as
\begin{widetext}
\begin{equation}
    \begin{aligned}
        \mathcal{J}_{nm}^{J}
        &=
        2 \frac{A}{(2\pi)^2}
        (T_{nm;0}^2 + \mathbf{T}_{nm} \cdot \mathbf{T}_{nm})
        \int \! \mathrm{d}^2 \mathbf{k} \, \mathrm{d}^2 \mathbf{k}' \,
        u(E_{n, \mathbf{k}}, E_{m, \mathbf{k}'}; \beta)
        K_n(k)K_m^*(k')
        \hat{\mathbf{d}}_n(\hat{\mathbf{k}}) \cdot \hat{\mathbf{d}}_m^*(\hat{\mathbf{k}}') ,
        \\
        \mathcal{J}_{nm}^{\mathrm{DM}}
        &=
        4i \frac{A}{(2\pi)^2}
        (T_{nm;0} \mathbf{T}_{nm})
        \int \! \mathrm{d}^2 \mathbf{k} \, \mathrm{d}^2 \mathbf{k}' \,
        u(E_{n, \mathbf{k}}, E_{m, \mathbf{k}'}; \beta)
        K_n(k)K_m^*(k')
        \hat{\mathbf{d}}_n(\hat{\mathbf{k}}) \times \hat{\mathbf{d}}_m^*(\hat{\mathbf{k}}'),
        \\
        \mathcal{J}_{nm}^\Gamma
        &=
        -4 \frac{A}{(2\pi)^2}
        {T}^a_{nm} {T}^b_{nm}
        \int \! \mathrm{d}^2 \mathbf{k} \, \mathrm{d}^2 \mathbf{k}' \,
        u(E_{n, \mathbf{k}}, E_{m, \mathbf{k}'}; \beta)
        K_n(k)K_m^*(k)'
        \hat{{d}}_n^a(\hat{\mathbf{k}}) \hat{{d}}_m^{b*}(\hat{\mathbf{k}}'),
    \end{aligned}
    \label{coupling_integrals.coupled_spin_orbit}
\end{equation}
where $A$ is the area of the two-dimensional superconducting grain.
Above, the weighting factor is given by
\begin{equation}
    u(E_{n, \mathbf{k}}, E_{m, \mathbf{k}'}; \beta) = \frac{\Delta_0^2 h^2(\mathbf{k}, \mathbf{k}')}{2E_{n, \mathbf{k}} E_{m, \mathbf{k}'}}
    \left (
    \frac{n_F(E_{n, \mathbf{k}}) + n_F(E_{m, \mathbf{k}'}) - 1}{E_{n, \mathbf{k}}+ E_{m, \mathbf{k}'}}
    -
    \frac{n_F(E_{n, \mathbf{k}}) - n_F(E_{m, \mathbf{k}'})}{E_{n, \mathbf{k}} - E_{m, \mathbf{k}'}}
    \right ),
\end{equation}
\end{widetext}
and
at zero temperature, this reduces to
\begin{equation}
    \left. u(E_{n, \mathbf{k}}, E_{m, \mathbf{k}'})\right|_{T=0} 
    = - \frac{\Delta_0^2 h^2(\mathbf{k}, \mathbf{k}')}{2E_{n, \mathbf{k}} E_{m, \mathbf{k}'}
    (E_{n, \mathbf{k}}+ E_{m, \mathbf{k}'})}.
\end{equation}
For example, when $h(\mathbf{k}, \mathbf{k}')$
is maximally weighted when $\mathbf{k} = \mathbf{k}'$, 
though $\hat{\mathbf{d}}(\hat{\mathbf{k}})$ is an odd function of $\hat{\mathbf{k}}$, the integrals in Eq.~\eqref{coupling_integrals.coupled_spin_orbit} can be nonvanishing.



\section{Microscopic models and effective tunneling in the presence of frustrated spin textures}

\label{sec:models_and_tunneling}

From the microscopic model of the Josephson coupling, 
two essential ingredients emerge for realizing the general Josephson free energy in 
Eq.~\eqref{triplet_triplet_josephson_form_factor.condensed}:
spin-dependent Josephson tunneling processes satisfying Eq.~\eqref{dm_coupling_condition}
and
nonvanishing triplet pairing correlations.
These two conditions are naturally realized in, for example, a system with itinerant electrons coupled to a frustrated local exchange field.
In this section, we focus on the first condition and demonstrate that the $s$-$d$ exchange can give rise to an effective spin-dependent tunneling that is sensitive to the frustration of the underlying local spin moments.
Later, in Sec.~\ref{sec:pairing_correlations}, we demonstrate the spin triplet pairing correlations.

\subsection{Three-sublattice \texorpdfstring{$s$-$d$}{s-d} model on geometrically frustrated lattice}

\label{subsec:three-sublattice_models}

We consider a minimal $s$-$d$ model on a geometrically frustrated system, such as a three-sublattice model on a kagome or triangular lattice.
The system consists of itinerant $s$ electrons coupled to frustrated localized spins, which act as a local exchange field.
The Hamiltonian for the $s$-$d$ system in the normal state is given by
\begin{equation}
    H
    = H_\mathrm{kin} + H_{sd},
    \label{hamiltonian_sd}
\end{equation}
in which $H_\mathrm{kin}$ and $H_{sd}$ describe nearest neighbour hopping and local $s$-$d$ exchange.
The kinetic term is
\begin{equation}
    H_{\mathrm{kin}} = t_0 \sum_{\langle \mathbf{r}_i, \mathbf{r}_j \rangle, \alpha}
    c^\dagger_{\mathbf{r}_i, \alpha}
    c_{\mathbf{r}_j, \alpha}
    -\mu \sum_{\mathbf{r}_i, \alpha} c^\dagger_{\mathbf{r}_i, \alpha} c_{\mathbf{r}_i, \alpha},
    \label{kinetic.NN.real_space}
\end{equation}
in which $t_0<0$ is the hopping amplitude, $\mu$ the chemical potential, and $c^\dagger_{\mathbf{r}_i, \alpha}$ the creation operator for $s$ electrons at site $\mathbf{r}_i$ and spin $\alpha = {\uparrow, \downarrow}$.
The $s$-$d$ coupling, at the mean-field level, is given by~\cite{Anderson1961, Kondo1962, Kondo1964, Schrieffer1966, Shiba1968, Chen2014, Coleman2015}
\begin{equation}
    H_{sd} = 
    J_{sd}
    \sum_{\mathbf{r}_j, \alpha, \alpha'}
    c^{\dagger}_{\mathbf{r}_j, \alpha}
    [
    \boldsymbol{\sigma}_{\alpha, \alpha'}
    \cdot
    \mathbf{s}_j
    ]
    c_{\mathbf{r}_j, \alpha'},
    \label{sd_coupling_ham.real_space}
\end{equation}
with
$J_{sd}$ being the amplitude of the $s$-$d$ coupling.
The local exchange field is described by the spin moment ${\mathbf{s}}_j = \langle d^\dagger_{\mathbf{r}_j, \alpha} \boldsymbol{\sigma}_{\alpha, \alpha'} d_{\mathbf{r}_j, \alpha'} \rangle/2$, with $d^\dagger_{\mathbf{r}_j, \alpha}$ being the creation operator for $d$ electrons.
In this work, we treat the local spins as static classical fields.

\begin{figure}
    \centering
    \includegraphics[width=\linewidth]{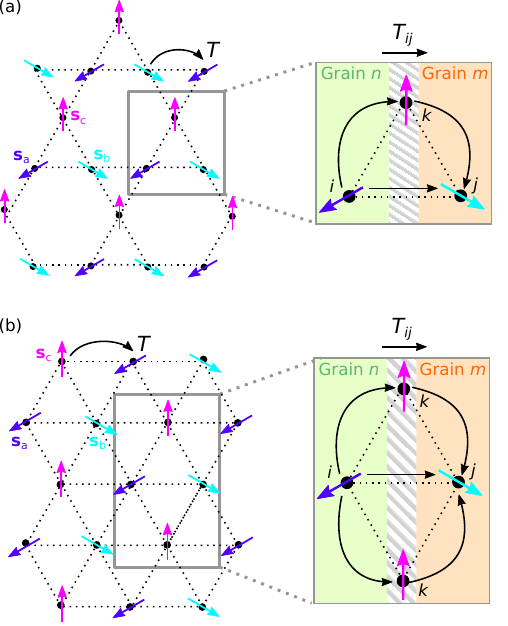}
    \caption{Local spin moments and effective tunneling for the three-sublattice systems for the (a)~kagome and (b)~triangular lattices.
    Nearest neighbour hopping between sites $i$ and $j$ on the boundaries of superconducting grains $n$ and $m$ consists of nearest neighbour hopping and effective tunneling from the underlying spin texture.
    In addition, there are higher order tunneling processes mediated by third spin $\mathbf{s}_k$.
    }
    \label{fig:tunneling}
\end{figure}

The three magnetic sublattices are labelled $a$, $b$, and $c$, as shown in Fig.~\ref{fig:tunneling}.
The local spin moment at site $j$ is given by $\mathbf{s}_j = \mathbf{s}_a$, $\mathbf{s}_b$, or $\mathbf{s}_c$, depending on the corresponding magnetic sublattice, and
the geometrically frustrated lattice leads to a noncollinear spin texture.
We consider the three spins comprising the three-sublattice system being in a $120^\circ$ ordered state with out-of-plane canting,
\begin{align}
    \hat{\mathbf{s}}_a
    &= \nonumber
    (\cos\theta_0 \cos\varphi_0, 
    \cos\theta_0 \sin\varphi_0, 
    \sin\theta_0
    )^\mathrm{T}, 
    \\
    \hat{\mathbf{s}}_b 
    &= \nonumber
    (\cos\theta_0 \cos(\varphi_0+\nu \frac{2\pi}{3}), 
    \cos\theta_0 \sin(\varphi_0+\nu \frac{2\pi}{3}), 
    \sin\theta_0
    )^\mathrm{T},
    \\
    \hat{\mathbf{s}}_c &= 
    (\cos\theta_0 \cos(\varphi_0-\nu \frac{2\pi}{3}), 
    \cos\theta_0 \sin(\varphi_0-\nu \frac{2\pi}{3}), 
    \sin\theta_0
    )^\mathrm{T}.
    \label{three_spins}
\end{align}
Here, $\varphi_0$ is a constant, $\theta_0$ describes the out-of-plane canting, and $\nu = \pm 1$ is the sign of the vector chirality for the local spins.
In the ground state, the spins form a $120^\circ$ antiferromagnetic order ($\theta_0 = 0$).
When the local spins form a noncoplanar configuration ($\theta_0 \neq 0$), the scalar spin chirality, $\chi_{abc} = \mathbf{s}_a \cdot (\mathbf{s}_b \times \mathbf{s}_c)$, is finite.

\subsubsection{Kagome lattice}

Consider the system on the kagome lattice, shown schematically in Fig.~\ref{fig:tunneling}(a).
Nearest neighbour vectors are given by
$\boldsymbol{\delta}_1 = \frac{a}{2} (1, 0)^\mathrm{T}$,
$
\boldsymbol{\delta}_2 = \frac{a}{2} ( \frac{1}{2}, \frac{\sqrt{3}}{2})^\mathrm{T}$,
and
$
\boldsymbol{\delta}_3 = \boldsymbol{\delta}_2 - \boldsymbol{\delta}_1$,
and primitive lattice vectors are $\mathbf{a}_1 =  a (1, 0)^\mathrm{T}$ and
$\mathbf{a}_2 
= 
a ( \frac{1}{2}, \frac{\sqrt{3}}{2})^\mathrm{T}$.
The corresponding reciprocal lattice vectors satisfying $\mathbf{a}_i \cdot \mathbf{b}_j = 2\pi \delta_{ij}$ are given by
\begin{math}
    \mathbf{b}_1 = b ( \frac{\sqrt{3}}{2}, - \frac{1}{2} )^\mathrm{T}
\end{math}
and
\begin{math}
    \mathbf{b}_2 = b (0, 1)^\mathrm{T},
\end{math}
with $b = 4\pi/ \sqrt{3} a$.
In the $(\mathbf{c}_{a, \mathbf{k}}, \mathbf{c}_{b, \mathbf{k}}, \mathbf{c}_{c, \mathbf{k}})^\mathrm{T}$ basis, the Hamiltonian describing spin-independent nearest neighbour hopping is
\begin{equation}
    \mathcal{H}_\mathrm{kin}(\mathbf{k})
    =
    -
    \mu \mathbbm{1}_3 \otimes \sigma^0
    + 2 t_0 \left(
    \begin{array}{ccc}
         0 & \cos \alpha_1  & \cos \alpha_2
         \\
         \cos \alpha_1 & 0 & \cos \alpha_3
         \\
         \cos \alpha_2 & \cos \alpha_3 & 0
    \end{array}
    \right)
    \otimes \sigma^0,
    \label{kagome.kinetic}
\end{equation}
in which $\alpha_i \equiv \mathbf{k} \cdot \boldsymbol{\delta}_{i},$ with $i = 1,2,3$.
Here, $\mathbf{c}_{\mathbf{k}, i} = (c_{\mathbf{k}, i, \uparrow}, c_{\mathbf{k}, i, \downarrow})$ is the annihilation operator for electron spinor with momentum $\mathbf{k}$ and sublattice index $i = a,b,c$.
For only nearest neighbour hopping, the system features two dispersive bands and a flat band from the destructive interference, characteristic of the kagome geometry.
The $s$-$d$ exchange is local and diagonal in sublattice space,
\begin{equation}
    \mathcal{H}_{sd} = 
    J_{sd}
    \left(
    \begin{array}{ccc}
         \mathbf{s}_a \cdot \boldsymbol{\sigma} & 0 & 0
         \\
         0 & \mathbf{s}_b \cdot \boldsymbol{\sigma} & 0
         \\
         0 & 0 & \mathbf{s}_c \cdot \boldsymbol{\sigma}
    \end{array}
    \right).
    \label{sd_term}
\end{equation}
The system breaks time reversal symmetry, yet respects inversion symmetry with respect to the hexagon center or a lattice site.

Fermi surfaces for the $s$-$d$ model on the kagome lattice are shown in Fig.~\ref{fig:FS_spin_texture}(a).
For coplanar spin configuration, the spins of the itinerant electrons are in-plane.
For the given system parameters, there are two spin-split Fermi surfaces centered about the $K$ points, each displaying a helical spin texture from the $s$-$d$ coupling.
The winding numbers of Fermi surfaces about the $K$ points are $+1$ and $-1$.
The system can be viewed as displaying spin-valley locking arising from $s$-$d$ exchange~\cite{Zhang2026}.
Overall, the system preserves inversion symmetry, with states at $\mathbf{k}$ and $-\mathbf{k}$ having the same spin polarization, as discussed in Appendix~\ref{appendix:spin_texture_and_pairing_correlations}.

\subsubsection{Triangular lattice}

\label{subsec:triangular}

Next, we consider the system on a triangular lattice, as shown in Fig.~\ref{fig:tunneling}(b).
Nearest neighbour vectors are given by
$ \boldsymbol{\delta}_1 = \frac{a}{\sqrt{3}} (1, 0)^\mathrm{T}$,
$ \boldsymbol{\delta}_2 = \frac{ a}{\sqrt{3}} ( \frac{1}{2}, \frac{\sqrt{3}}{2})^\mathrm{T}$,
and
$\boldsymbol{\delta}_3 = 
\boldsymbol{\delta}_2 - \boldsymbol{\delta}_1$.
Due to the three-sublattice magnetic ordering, the unit cell is enlarged, with lattice vectors given by
$\mathbf{a}_1 = a ( \frac{\sqrt{3}}{2}, \frac{1}{2} )^\mathrm{T}$ and
$\mathbf{a}_2 = a ( 0, 1)^\mathrm{T}$.
Reciprocal lattice vectors  are given by
\begin{math}
    \mathbf{b}_1 = b (1, 0)^\mathrm{T}
\end{math}
and
\begin{math}
    \mathbf{b}_2 = b (- \frac{1}{2},  \frac{\sqrt{3}}{2} )^\mathrm{T},
\end{math}
with $b = 4\pi/ \sqrt{3} a$.
In the $(\mathbf{c}_{a, \mathbf{k}}, \mathbf{c}_{b, \mathbf{k}}, \mathbf{c}_{c, \mathbf{k}})^\mathrm{T}$ basis, the kinetic Hamiltonian is given by
\begin{equation}
    \mathcal{H}_\mathrm{kin}(\mathbf{k})
    =
    -\mu \mathbbm{1}_3 \otimes \sigma^0
    + t_0
    \left(
    \begin{array}{ccc}
        0 & 
        A(\mathbf{k})
        &
        A(-\mathbf{k})
        \\
        A(-\mathbf{k}) & 0 & A(\mathbf{k})
        \\
        A(\mathbf{k}) & A(-\mathbf{k}) & 0
    \end{array}
    \right) \otimes \sigma^0,
    \label{triangular_lattice.kinetic_ham}
\end{equation}
in which $A(\mathbf{k}) = ( e^{-i \mathbf{k} \cdot \boldsymbol{\delta}_1} + e^{i \mathbf{k} \cdot \boldsymbol{\delta}_2} + e^{-i \mathbf{k} \cdot \boldsymbol{\delta}_3})$.
Lastly, the $s$-$d$ term is the same as that in Eq.~\eqref{sd_term},
which is diagonal in sublattice space.
The system breaks both inversion and time reversal symmetries.

Furthermore, we also account for Ising-type spin-orbit coupling~\cite{Zhou2016} for the system on the triangular lattice, originating from the lack of inversion symmetry in the bulk system.
In the $(\mathbf{c}_{a, \mathbf{k}}, \mathbf{c}_{b, \mathbf{k}}, \mathbf{c}_{c, \mathbf{k}})^\mathrm{T}$ basis, the Ising spin-orbit coupling is given by
\begin{equation}
    \mathcal{H}_\mathrm{SO}(\mathbf{k})
    =
    \lambda_\mathrm{SO}
    \left(
    \begin{array}{ccc}
        0 & 
        i A(\mathbf{k})
        &
        -i A(-\mathbf{k})
        \\
        -i A(-\mathbf{k}) & 0 & i A(\mathbf{k})
        \\
        i A(\mathbf{k}) & -i A(-\mathbf{k}) & 0
    \end{array}
    \right)\otimes \sigma^z,
    \label{triangular_lattice.Ising_SO}
\end{equation}
in which $\lambda_\mathrm{SO}$ is the strength of the spin-orbit coupling.
Ising spin-orbit coupling has been shown to lead to spin-valley polarization in transition metal dichalcogenides.

\begin{figure}
    \centering
    \includegraphics[width= \linewidth]{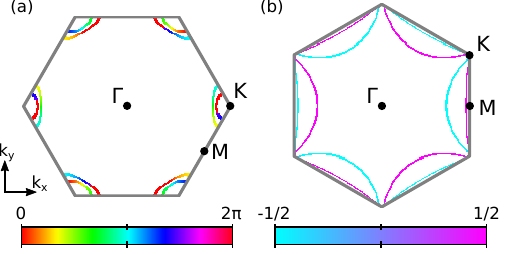}
    \caption{
    \textbf{(a)}~Fermi surfaces for the $s$-$d$ model on the kagome lattice, shown in the first Brillouin zone.
    States are plotted according to the azimuthal angle of their in-plane spin, $\arctan(\langle s_y \rangle/ \langle s_x \rangle)$.
    Parameters are 
    $\mu = -1.6 |t_0|$ and $J_{sd} = 0.2 |t_0|$.
    \textbf{(b)}~Fermi surfaces for the $s$-$d$ model on the triangular lattice, shown in the first Brillouin zone.
    States are colored on a continuum according to $z$-component of spin, $\langle s_z \rangle$.
    Parameters are 
    $\mu = -0.1 |t_0|$, $J_{sd} = 0.1 |t_0|$, and $\lambda_\mathrm{SO} = 0.3 |t_0|$. 
    For both the kagome and triangular lattices, the local spins $\mathbf{s}_a$, $\mathbf{s}_b$, and $\mathbf{s}_c$ form a coplanar $120^\circ$ ordered antiferromagnetic configuration corresponding to $\theta_0 = 0$, $\varphi_0 = 0$, and $\nu = -1$ in Eq.~\eqref{three_spins}.
    }
    \label{fig:FS_spin_texture}
\end{figure}

We show the Fermi surface for the $s$-$d$ model with Ising spin-orbit coupling in Fig.~\ref{fig:FS_spin_texture}(b).
With $s$-$d$ exchange, Ising spin-orbit coupling, or a combination of the two, there are spin-polarized Fermi surfaces, with the spin of itinerant electrons polarized in the $z$-direction.
Even for coplanar spin configuration, the $s$-$d$ exchange can lead to spins of the itinerant electrons polarized out-of-plane due to the geometric frustration, in contrast to the system on the kagome lattice, in which the spins lie in-plane.
Consequently, the $s$-$d$ coupling
can enhance the spin-valley locking in an Ising superconductor, depending on the sign of the vector chirality of the underlying spin configuration.

\subsection{Effective tunneling in the presence of \texorpdfstring{$s$-$d$}{s-d} exchange}

\label{subsec:effective_tunneling}

We next derive the effective tunneling of itinerant electrons in the presence of a local exchange field comprised of frustrated spin moments.
Consider the Hamiltonian $\mathcal{H} = \mathcal{H}_0 + \Sigma$.
Here, $\mathcal{H}_0$ denotes the Hamiltonian of the unperturbed system and refers to nearest neighbour hopping $\mathcal{H}_\mathrm{kin}$, and $\Sigma$ is the perturbation, corresponding to $s$-$d$ exchange $\mathcal{H}_{sd}$ in the context of this work.
The corresponding Green's function 
is given by
\begin{math}
    \mathcal{G}(i\omega) = (i\omega - \mathcal{H})^{-1}.
\end{math}
Let $\mathcal{G}_0(i\omega) = (i\omega - \mathcal{H}_0)^{-1}$ be the Green's function of the unperturbed Hamiltonian.
From Dyson's equation, it follows that
\begin{math}
    \mathcal{G}(i\omega) = (i\omega - \mathcal{H}_0 - \Sigma)^{-1}
    =
    \mathcal{G}_0(i\omega) + \mathcal{G}_0(i\omega) \Sigma  \mathcal{G}(i\omega),
\end{math}
or equivalently,~\cite{Abrikosov1975, Balatsky2006}
\begin{equation}
    \mathcal{G}(i\omega) = \mathcal{G}_0(i\omega) + \mathcal{G}_0(i\omega) \tilde{\Sigma}(i\omega) \mathcal{G}_0(i\omega),
\end{equation}
with $\tilde{\Sigma}$ serving as the $T$-matrix, defined recursively as
\begin{equation}
    \tilde{\Sigma}(i\omega) = \Sigma + \Sigma \mathcal{G}_0(i\omega)\tilde{\Sigma}(i\omega).
\end{equation}
In this work, we treat $s$-$d$ exchange $\mathcal{H}_{sd}$ perturbatively and retain terms up to third order in $J_{sd}$ to capture any nonvanishing spin chirality of the three-sublattice magnetic ordering.
From the above expansion, the effective $s$-$d$ exchange is given by
\begin{align}
    \tilde{\Sigma}_\mathrm{eff}(i\omega)
    \approx \ \nonumber
    &\mathcal{H}_{sd}
    +
    \mathcal{H}_{sd} \mathcal{G}_\mathrm{kin} (i\omega)
    \mathcal{H}_{sd}
    \\
    &+
    \mathcal{H}_{sd}
    \mathcal{G}_\mathrm{kin}(i\omega)
    \mathcal{H}_{sd}
    \mathcal{G}_\mathrm{kin} (i\omega)
    \mathcal{H}_{sd}.
    \label{sd_T_matrix_expansion}
\end{align}
Above, $\mathcal{G}_\mathrm{kin}(i\omega) = (i\omega - \mathcal{H}_\mathrm{kin})^{-1}$ is the free Green's function in the absence of $s$-$d$ exchange, which includes only spin-independent nearest neighbour hopping.

Now, we consider the off-diagonal matrix elements describing the effective tunneling processes between nearest neighbouring sites.
Considering the spin-independent nearest neighbour hopping and the $T$-matrix expansion of the $s$-$d$ exchange in
Eq.~\eqref{sd_T_matrix_expansion}, the effective tunneling in the spin-$1/2$ basis between nearest neighbouring sites $i$ and $j$ is given by
\begin{align}
    &\nonumber
    T_{ij}(i\omega)
    \equiv \langle i | (\mathcal{H}_0 
    + \tilde{\Sigma}_\mathrm{eff})| j \rangle
    \\
    &= \nonumber
    t_0 \sigma^0
    +
    J_{sd}^2
    \Big(
    \left(\mathbf{s}_{i} \cdot \boldsymbol{\sigma} \right)
    [\mathcal{G}_\mathrm{kin}(
    i\omega)]_{i j}
    \left(\mathbf{s}_{j} \cdot \boldsymbol{\sigma} \right)
    \Big)
    \\
    &+
    J_{sd}^3 
    \sum_{k}
    \Big (
    \left(\mathbf{s}_{i} \cdot \boldsymbol{\sigma} \right)
    [\mathcal{G}_\mathrm{kin}( 
    i\omega)]_{i,k}
    \left(\mathbf{s}_{k} \cdot \boldsymbol{\sigma} \right)
    [\mathcal{G}_\mathrm{kin}(
    i\omega)]_{k,j}
    \left(\mathbf{s}_{j} \cdot \boldsymbol{\sigma} \right)
    \Big).
    \label{Tij_third_order_sd}
\end{align}
Above, $\mathbf{s}_{i}$ is the local spin moment at site $i$, and the summation is taken over sites $k$ which are nearest neighbours to both sites $i$ and $j$.
In Figs.~\ref{fig:tunneling}(a) and (b), we show schematically the effective tunneling in the presence of the local spin moments for the kagome and triangular lattices, respectively.

Using the fact that $\mathcal{G}_\mathrm{kin}(i\omega)$ is spin-independent and that the $s$-$d$ exchange is on-site,
the effective tunneling matrix reduces to
\begin{align}
    T_{i j}(i\omega)
    = 
    & \nonumber
    \ t_0 
    \sigma^0
    +
    J_{sd}^2
    [\mathcal{G}_\mathrm{kin}(i\omega)]_{i j}
    \alpha_{ij}
    \sigma^0
    \\
    & + \nonumber
    i J_{sd}^2
    [\mathcal{G}_\mathrm{kin}(i\omega)]_{i j}
    (\boldsymbol{\beta}_{ij} \cdot
    \boldsymbol{\sigma})
    \\
    &- \nonumber
    i J_{sd}^3
    \sum_{k}
    [\mathcal{G}_\mathrm{kin}(i\omega)]_{i k}
    [\mathcal{G}_\mathrm{kin}(i\omega)]_{k j}
    \chi_{ijk}
    \sigma^0
    \\
    &+
    J_{sd}^3
    \sum_{k}
    [\mathcal{G}_\mathrm{kin}(i\omega)]_{i k}
    [\mathcal{G}_\mathrm{kin}(i\omega)]_{k j}
    \left(
    \boldsymbol{\gamma}_{ijk}
    \cdot
    \boldsymbol{\sigma}
    \right).
    \label{effective_tunneling_Jsd_expansion}
\end{align}
Above, we have defined the factors
\begin{equation}
    \begin{gathered}
        \alpha_{ij} \equiv \mathbf{s}_{i} \cdot \mathbf{s}_{j};
        \hspace{2em}
        \boldsymbol{\beta}_{ij} \equiv \mathbf{s}_{i} \times \mathbf{s}_{j};
        \\
        \chi_{ijk} \equiv \mathbf{s}_{i} \cdot (\mathbf{s}_{j}\times \mathbf{s}_{k});
        \\
        \boldsymbol{\gamma}_{ijk} \equiv
        (\mathbf{s}_{i} \cdot \mathbf{s}_{k})\mathbf{s}_{j}
        -
        (\mathbf{s}_{i} \cdot \mathbf{s}_{j})\mathbf{s}_{k}
        +
        (\mathbf{s}_{j} \cdot \mathbf{s}_{k})\mathbf{s}_{i},
    \end{gathered}
    \label{spin_factors}
\end{equation}
which are dependent on the underlying local exchange field.
For the three-sublattice system, spins $\mathbf{s}_i$, $\mathbf{s}_j$, and $\mathbf{s}_k$ correspond to the three spins $\mathbf{s}_a$, $\mathbf{s}_b$, and $\mathbf{s}_c$.
The first term, $\alpha_{ij}$, is even under time reversal symmetry and under inversion with respect to the center of the bond between sites $i$ and $j$, and it is maximized for a collinear spin configuration.
$\boldsymbol{\beta}_{ij}$ is likewise even under time reversal symmetry, but it is odd under the inversion above and maximized for noncollinear spin configurations.
The scalar spin chirality $\chi_{ijk}$ breaks both inversion and time reversal symmetries.
Lastly, the higher order term $\boldsymbol{\gamma}_{ijk}$ breaks time reversal symmetry but preserves inversion symmetry.
The three terms $\boldsymbol{\beta}_{ij}$, $\boldsymbol{\gamma}_{ijk}$, and $\chi_{ijk}$ can be nonvanishing for noncollinear spin textures.

In principle, finite scalar spin chirality can also induce an emergent gauge field via Peierls substitution, which would enter into the free Green's function $\mathcal{G}_\mathrm{kin}$\cite{Peierls1933, Hatsugai1993, Ohgushi2000}.
This mechanism has been shown, for example, to generate an anomalous Hall effect~\cite{Ohgushi2000, Xiao2010} in addition to stabilizing superconductivity when pairing occurs between bands with opposite gauge charges~\cite{Dong2024}.
In this work, we neglect the Peierls phase as we are primarily interested in the spin-dependent tunneling factors arising in the presence of a local exchange field.
Moreover, the formalism in this work, namely the anisotropic Josephson couplings in Eq.~\eqref{free_energy_josephson_coupling_discrete}, can persist in coplanar spin configurations without scalar spin chirality.
Lastly, we discuss the effective tunneling in the limit of strong $J_{sd}$ in Appendix~\ref{appendix:strong_Jsd_tunneling}.

\section{Spin triplet pairing correlations arising from \texorpdfstring{$s$-$d$}{s-d} exchange or spin-orbit coupling}

\label{sec:pairing_correlations}

We now discuss the spin triplet correlations
in the three-sublattice systems.
We consider two representative cases: the case of only $s$-$d$ exchange without spin-orbit coupling, and the case of 
spin-orbit coupling.
In general, spin-orbit coupling can give rise to an admixture of spin singlet and spin triplet pairing correlations, owing to broken inversion symmetry~\cite{Gorkov2001, Smidman2017, Amundsen2024}, and here, we focus on the role of Ising spin-orbit coupling as a representative example.
The following analysis of the bulk pairing order 
neglects the effects of Josephson coupling, which are discussed in Sec.~\ref{sec:josephson_three_sublattice}.

The BdG
Hamiltonian kernel in the Nambu basis $(\mathbf{c}_{\mathbf{k}}, \mathbf{c}^\dagger_{-\mathbf{k}})^\mathrm{T}$ takes the form
\begin{equation}
    \mathcal{H}_\mathrm{BdG}(\mathbf{k})
    =
    \left(
    \begin{array}{cc}
         \mathcal{H}
         (\mathbf{k})
         & \Delta(\mathbf{k})
         \\
         \Delta^\dagger(\mathbf{k})
         &
         - \mathcal{H}^\mathrm{T}(-\mathbf{k})
    \end{array}
    \right),
    \label{BdG_Ham}
\end{equation}
in which $\mathbf{c}_\mathbf{k} = (\mathbf{c}_{a, \mathbf{k}}, \mathbf{c}_{b, \mathbf{k}}, \mathbf{c}_{c, \mathbf{k}})$.
Above, $\mathcal{H}(\mathbf{k})$ is the kernel of the Hamiltonian in Eq.~\eqref{hamiltonian_sd}, and $\Delta(\mathbf{k})$ is the pairing gap function, which can be either intrinsic or proximitized.
The anomalous Green's function, $\mathcal{F}_{i, \alpha; j, \beta} (\mathbf{k}; i\omega) = - \int \mathrm{d} \tau  e^{i\omega \tau} \langle \mathcal{T}_\tau c_{-\mathbf{k}, j , \beta} (\tau) c_{\mathbf{k}, i, \alpha}(0) \rangle$, encodes the superconducting pairing correlations and corresponds to the off-diagonal component of the BdG Green's function $\mathcal{G}_\mathrm{BdG}(\mathbf{k}; i\omega) = (i\omega - \mathcal{H}_\mathrm{BdG}(\mathbf{k}))^{-1}$~\cite{Abrikosov1975, Lueders1971}.
The anomalous Green's function
is given by
\begin{align}
    \mathcal{F}(\mathbf{k}; i\omega) 
    = 
    - \mathcal{G}
    (\mathbf{k}; i\omega)
    \Delta(\mathbf{k})
    D(\mathbf{k}; i\omega)
    \mathcal{G}^\mathrm{T}(-\mathbf{k}; -i\omega),
    \label{anomalous_GF.freq_space}
\end{align}
in which 
\begin{math}
    \mathcal{G}
    (\mathbf{k}; i\omega) 
    =
    (i\omega - \mathcal{H}
    (\mathbf{k}))^{-1}
\end{math}
is the normal single-particle Green's function, including
$s$-$d$ exchange
and spin-orbit coupling, and 
\begin{math}
    D(\mathbf{k}; i\omega) = [
    \mathbbm{1} + \mathcal{G}^\mathrm{T}(-\mathbf{k}; -i\omega)
    \Delta^\dagger (\mathbf{k}) \mathcal{G}
    (\mathbf{k}; i\omega) \Delta(\mathbf{k}) ]^{-1}.
\end{math}
In the following, we consider the case of momentum-independent, intrasublattice proximitized $s$-wave pairing,
$\Delta (\mathbf{k}) = \Delta_0 \mathbbm{1}_{3\times 3} \otimes (i\sigma^y)$, 
with $\Delta_0$ being the pairing amplitude.

\subsection{Pairing correlations in \texorpdfstring{$s$}{s}-\texorpdfstring{$d$}{d} model}

\label{subsec:pairing_correlations.sd_model}

We first analyze pairing correlations arising solely from spin-independent hopping and $s$-$d$ exchange.
Coupling to the exchange field formed by local spin moments naturally generates spin triplet components in the pairing correlations.
To demonstrate this effect, we expand perturbatively in the $s$-$d$ coupling $J_{sd}$.
The single-particle Green's function can be expressed via a Dyson series as
\begin{equation}
    \mathcal{G}
    (\mathbf{k}; i\omega)
    =
    \mathcal{G}_\mathrm{kin}(\mathbf{k}; i\omega)
    \sum_{n \geq 0}
    [\mathcal{H}_{sd} \mathcal{G}_\mathrm{kin}(\mathbf{k}; i\omega)]^n,
    \label{Gband.dyson_expansion}
\end{equation}
with $\mathcal{G}_\mathrm{kin}(\mathbf{k}; i\omega) = (i\omega - \mathcal{H}_\mathrm{kin}(\mathbf{k}))^{-1}$ being the free Green's function in the absence of the $s$-$d$ exchange.
In the linearized gap regime, with $|\Delta_0| < |J_{sd}|$ and $|\Delta_0| \ll |t_0|$, the anomalous Green's function
can be systematically expanded in orders of $J_{sd}$ as $\mathcal{F}(\mathbf{k}; i\omega) = \sum_N \mathcal{O}(J_{sd}^N)$, in which the $N\mathrm{th}$ order contribution is given by
\begin{align}
    & 
    \nonumber 
    \mathcal{O}(J_{sd}^{N})
    = 
    \Delta_0
    \sum_{n = 0}^{N}
    (-1)^{N-n+1}
    \mathcal{G}_\mathrm{kin}(\mathbf{k}; i\omega) 
    \Big[
    \mathcal{H}_{sd} \mathcal{G}_\mathrm{kin}(\mathbf{k}; i\omega)
    \Big]^n
    \\
    &
    \times
    \mathcal{G}_\mathrm{kin}^\mathrm{T}(-\mathbf{k}; -i\omega)
    \Big[
    \mathcal{H}_{sd} \mathcal{G}_\mathrm{kin}^\mathrm{T}(-\mathbf{k}; -i\omega)
    \Big]^{N-n}
    (\mathbbm{1}_{3 \times 3} \otimes i\sigma^y).
\end{align}
Above, $\mathcal{G}_\mathrm{kin}(\mathbf{k}; i\omega)$ has off-diagonal components corresponding to intersublattice hopping while $\mathcal{H}_{sd}$ introduces spin-sublattice coupling via the local exchange field.
For noncollinear spin texture, this leads to emergence of mixed-parity superconducting correlations in both spin singlet and spin triplet channels.

To demonstrate explicitly,  
we decompose the anomalous Green's function between sublattices $i$ and $j$ in Eq.~\eqref{anomalous_GF.freq_space} as
\begin{equation}
    \mathcal{F}_{ij}(\mathbf{k}; i\omega) = 
    \Big[
    f_{0; ij}(\mathbf{k}; i\omega)
    +
    \mathbf{f}_{ij}(\mathbf{k}; i\omega) \cdot \boldsymbol{\sigma} 
    \Big]
    (i\sigma^y),
    \label{singlet_triplet_pairing_correlations}
\end{equation}
in which $f_{0; ij}$ and $\mathbf{f}_{ij}$ denote the singlet and triplet pairing correlations respectively.
The vector $\mathbf{f}_{ij}$ plays the role of the $d$ vector, corresponding to induced spin triplet correlations, and its direction and magnitude depend on the underlying spin texture, lattice geometry, and relative strength of $s$-$d$ coupling.

For the system on the kagome lattice, the noncollinear spin texture discussed in Sec.~\ref{sec:models_and_tunneling} breaks time reversal but preserves inversion symmetry with respect to the hexagon center.
As illustrated in Fig.~\ref{fig:FS_spin_texture}(a), the spins of the itinerant electrons at the Fermi surface are polarized in-plane with helical spin textures.
Due to the inversion symmetry, 
the spin of electrons at $\mathbf{k}$ and $-\mathbf{k}$ participating in the zero center-of-mass momentum pairing have the same spin.
Consequently, in the weak coupling regime, pairing correlations for states at the Fermi surface can be present, but are generally weak and do not open a gap,  as shown in Fig.~\ref{fig:pairing_correlations}(a) and 
discussed in Appendix~\ref{appendix:spin_texture_and_pairing_correlations}.
To realize stronger pairing correlations for states at the Fermi surface, one can consider a different pairing interaction channel or introduce additional symmetry-breaking terms.
For example, near the interface with an $s$-wave superconductor, inversion symmetry breaking at the junction interface can lead to spin triplet pairing correlations~\cite{Zhang2026}.


In contrast, for the system on the triangular lattice, $s$-$d$ exchange breaks both time reversal and inversion symmetries for noncollinear spin textures, allowing for an admixture of spin singlet and triplet superconducting pairing correlations for states at the Fermi level.
As shown in Fig.~\ref{fig:FS_spin_texture}(b), the $s$-$d$ coupling results in spin-valley polarization, with the spins of itinerant electrons with momentum $\mathbf{k}$ and $-\mathbf{k}$ being antiparallel,
which is compatible with the $s$-wave pairing gap function.
This results in a mixture of spin singlet pairing and spin triplet equal-spin pairing correlations, analogous to an Ising superconductor~\cite{Zhou2016}.
The pairing correlations are confirmed in Fig.~\ref{fig:pairing_correlations}(b), where both spin singlet and spin triplet pairing correlations for states at the Fermi surface are of similar order of magnitude.
While the relative magnitude and direction of $\mathbf{f}_{ij}(\mathbf{k})$ is largely dependent on the system parameters, including the underlying local exchange field and relative strength of $J_{sd}$, the general features---namely the coexistence of spin singlet and spin triplet superconducting pairing correlations---are robust over a wide range of parameters.

\begin{figure}
    \centering
    \includegraphics[width=\linewidth]{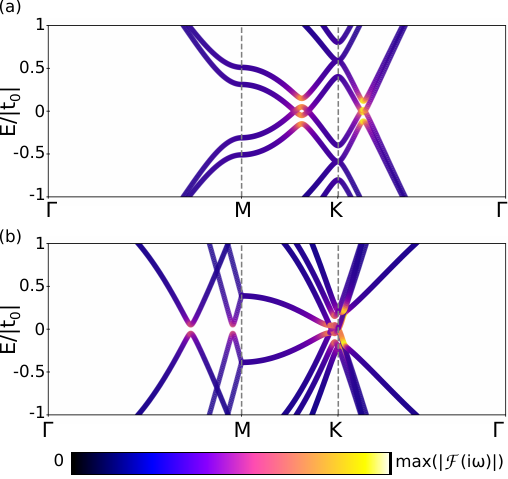}
    \caption{BdG excitation spectra and spin triplet pairing correlations for systems on the (a)~kagome lattice and (b)~triangular lattice.
    The spectrum is colored on a continuum according to the magnitude of the total spin triplet pairing correlations, including inter- and intrasublattice pairing.
    Parameters for the system are the same as those in Fig.~\ref{fig:FS_spin_texture}, and the magnitude of the pairing gap function is $\Delta_0 = 0.05 |t_0|$.
    }
    \label{fig:pairing_correlations}
\end{figure}

\subsection{Pairing correlations in an Ising superconductor}

In systems with strong Ising spin-orbit coupling, pairing correlations in the isolated bulk superconductor can primarily be driven by the spin-orbit coupling,
as in the case of an Ising superconductor~\cite{Zhou2016}.
While the magnitude of the $s$-$d$ exchange and the Ising spin-orbit coupling can be on the same energy scale, we focus exclusively on the role of the Ising spin-orbit coupling in the following for simplicity.
The addition of $s$-$d$ exchange can further enhance the spin-valley locking, as described in Sec.~\ref{subsec:triangular}.

We consider the band Hamiltonian $\mathcal{H}_\mathrm{band}(\mathbf{k})$ accounting for the nearest neighbour hopping and Ising spin-orbit coupling.
The band Hamiltonian for the triangular lattice is given by
\begin{align}
    \mathcal{H}_\mathrm{band}(\mathbf{k})
    &=
    \mathcal{H}_\mathrm{kin} (\mathbf{k}) + \mathcal{H}_\mathrm{SO} (\mathbf{k}),
    \label{band_ham_kin_SO_triangular}
\end{align}
in which $\mathcal{H}_\mathrm{kin}$ in Eq.~\eqref{triangular_lattice.kinetic_ham} describes the hopping and on-site chemical energy, and $\mathcal{H}_\mathrm{SO}$ in Eq.~\eqref{triangular_lattice.Ising_SO} describes the Ising spin-orbit coupling.
Following Ref.~\citenum{Zhou2016}, which studies Ising superconductivity for a two-band model, we consider proximitizing the three-sublattice system with $s$-wave superconductivity, with the pairing gap function taking the form $\Delta(\mathbf{k}) = {\Delta_0 \mathbbm{1}_{3\times 3} \otimes i\sigma^y}$.

Because the band Hamiltonian in Eq.~\eqref{band_ham_kin_SO_triangular} is diagonal in spin-space, the Green's function $\mathcal{G}_\mathrm{band}(\mathbf{k}; i\omega) = (i\omega - \mathcal{H}_\mathrm{band}(\mathbf{k}))^{-1}$ can be decomposed as
\begin{equation}
    \mathcal{G}_\mathrm{band}( \mathbf{k}; i\omega) = g_0( \mathbf{k}; i\omega) \otimes \sigma^0 + g_z( \mathbf{k}; i\omega) \otimes \sigma^z.
\end{equation}
Here, $g_0(\mathbf{k}; i\omega)$ and $g_z(\mathbf{k}; i\omega)$ are $3 \times 3$ matrices in sublattice space.
In the linearized gap regime, the anomalous Green's function takes the form $\mathcal{F}(\mathbf{k}; i\omega) \approx - \mathcal{G}_\mathrm{band}(\mathbf{k}; i\omega) \Delta(\mathbf{k}) \mathcal{G}_\mathrm{band}^\mathrm{T}(- \mathbf{k}; -i\omega)$ and can be decomposed into spin singlet and spin triplet components, $\mathcal{F}(\mathbf{k}; i\omega) = (f_0(\mathbf{k}; i\omega) + \mathbf{f}(\mathbf{k}; i\omega) \cdot \boldsymbol{\sigma})(i\sigma^y)$, in which
\begin{align}
    f_0
    &\approx \nonumber 
    \Delta_0 [ g_0(\mathbf{k}; i\omega) g_0(-\mathbf{k}; -i\omega)
    - g_z(\mathbf{k}; i\omega)g_z(-\mathbf{k}; -i\omega)
    ]
    \\
    \mathbf{f}
    &\approx 
    \Delta_0[
    g_0(-\mathbf{k}; -i\omega) g_z(\mathbf{k}; i\omega)
    -
    g_0(\mathbf{k}; i\omega)  g_z(-\mathbf{k}; -i\omega)
    ] \hat{z}.
\end{align}
Here, $\mathbf{f} = \mathbf{f}(\mathbf{k}; i\omega)$ and $f_0 = f_0(\mathbf{k}; i\omega)$ detail the intrasublattice and intersublattice pairing correlations, satisfying $[f_0(\mathbf{k}; i\omega)]_{ij} = [f_0(-\mathbf{k}; i\omega)]_{ji}$ and 
$[\mathbf{f}(\mathbf{k}; i\omega)]_{ij} = -[\mathbf{f}(-\mathbf{k}; i\omega)]_{ji}$, with $i$ and $j$ being the sublattice indices.
It follows that the spin triplet pairing components arise from the Ising spin-orbit coupling, with the $d$ vector polarized along the $z$ direction, corresponding to equal-spin pairing.

\section{Effective Josephson coupling in the presence of frustrated spin textures}

\label{sec:josephson_three_sublattice}

In Sec.~\ref{sec:models_and_tunneling}, we demonstrated that coupling to a local exchange field can give rise to an effective spin-dependent tunneling of itinerant electrons, and in Sec.~\ref{sec:pairing_correlations}, we showed that either $s$-$d$ exchange or spin-orbit coupling can induce spin triplet pairing correlations for the three-sublattice system.
In the following, we consider a system composed of many superconducting grains and demonstrate that the effective tunneling from $s$-$d$ exchange can generate anisotropic Josephson couplings between the spin triplet pairing correlations.
These couplings, in turn, can stabilize spatially varying $d$ vector textures.

\begin{figure*}
    \centering
    \includegraphics[width=\linewidth]{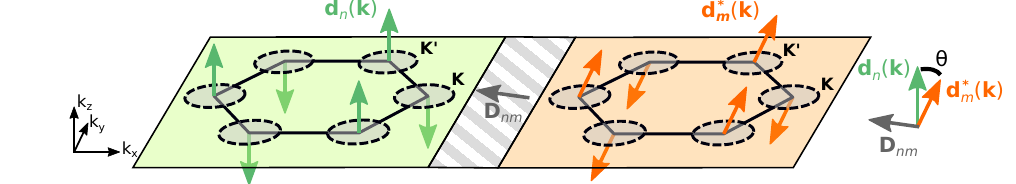}
    \caption{
    Spatially inhomogeneous $d$ vector textures arising from anisotropic Josephson coupling in a hybrid picture of real and momentum space.
    Depicted is a microscopic Josephson junction between superconducting grains $n$ and $m$ described by $d$ vectors $\mathbf{d}_n(\mathbf{k})$ and $\mathbf{d}_m(\mathbf{k})$ in a hybrid picture of momentum- and real-space.
    An example $d$ vector is shown in momentum space, with Fermi pockets about the $K$ points.
    Minimization with respect to the bulk free energy and Josephson coupling can lead to a spatially nonuniform $d$ vector configuration.
    }
    \label{fig:kspace_dvector_texture}
\end{figure*}

Consider the Josephson coupling between two adjacent superconducting grains $n$ and $m$, for which the spin triplet pairing correlations are described by $d$ vectors $\mathbf{d}_n(\mathbf{k})$ and $\mathbf{d}_m(\mathbf{k})$, as shown in Fig.~\ref{fig:kspace_dvector_texture}.
We assume that states at the grain boundary primarily contribute to the Josephson tunneling.
As such, the tunneling process will depend on the local spin configuration at the interface between two superconducting grains, with the total Josephson coupling obtained by summing over all microscopic tunneling processes along the grain boundary.
For simplicity,
we consider tunneling of the form $T_{nm}(\mathbf{k}, \mathbf{k}') = T_{nm} \delta_{\mathbf{k}, \mathbf{k}'}$.
From the nearest neighbour effective tunneling in Eq.~\eqref{effective_tunneling_Jsd_expansion}, the spin-independent and spin-dependent tunneling of itinerant electrons across the grain boundary are given by
\begin{math}
    T_{nm; 0}
    =
    \sum_{i \in \Sigma_n, j \in \Sigma_m}
    T_{ij; 0}
\end{math}
and
\begin{math}
    \mathbf{T}_{nm}
    =
    \sum_{i \in \Sigma_n, j \in \Sigma_m}
    \mathbf{T}_{ij}.
\end{math}
Here, summation of nearest neighbouring sites $i$ and $j$ is taken over the boundaries $\Sigma_n$ and $\Sigma_m$ of 
superconducting grains $n$ and $m$ respectively,
and $T_{ij}$ is the effective tunneling in Eq.~\eqref{effective_tunneling_Jsd_expansion}, which arises from $s$-$d$ exchange at the interface.

As an illustrative example, we examine the model on a triangular lattice, shown schematically in Fig.~\ref{fig:tunneling}(b).
For the three-sublattice system, nearest neighbours correspond to sublattices with spins $\mathbf{s}_{i}$ and $\mathbf{s}_{j}$, and the summation in the third order terms corresponds to the third sublattice, which we label as $\mathbf{s}_{k}$.
From Eq.~\eqref{effective_tunneling_Jsd_expansion}, the effective spin-independent and spin-dependent tunneling between nearest neighbours for the triangular lattice is given by
\begin{equation}
    \begin{aligned}
        T_{nm; 0}
        &\approx
        \underset{j \in \Sigma_m}{\sum_{i \in \Sigma_n,}}
        \bigg(
        t_0  {-} \frac{J_{sd}^2}{t_0} 
        \alpha_{ij}
        -
        2i \frac{J_{sd}^3}{t_{0}^2}
        \chi_{ijk}
        \bigg),
        \\
        \mathbf{T}_{nm}
        &\approx
        \underset{j \in \Sigma_m}{\sum_{i \in \Sigma_n,}}
        \bigg(
        {-}i \frac{J_{sd}^2}{t_0} 
        \boldsymbol{\beta}_{ij}
        +
        2 \frac{J_{sd}^3}{t_0^2}
        \boldsymbol{\gamma}_{ijk}
        \bigg).
    \end{aligned}
    \label{spin_dependent+independent_tunneling}
\end{equation}
Here, we approximate $[\mathcal{G}_\mathrm{kin}]_{i j} \approx - (1/t_0) \sigma^0$ for nearest neighbours $i$ and $j$, which is valid for states near the Fermi surface.
The index $k$ corresponds to the third sublattice mediating the higher order tunneling process between $i$ and $j$, with $i$, $j$, and $k$ corresponding to the three sublattices in the same plaquette.

For simplicity, let us consider a $d$ vector which is valley-polarized in momentum space, corresponding to the case in Sec.~\ref{subsec:josephson_coupling.decoupled_spin_orbital}.
As an example, we take $g_m(\mathbf{k})$ in Eq.~\eqref{decoupled_spin_orbital_anomalous_GF}
to have $f$-wave symmetry, as depicted in Fig.~\ref{fig:kspace_dvector_texture}.
From the form of the tunneling in Eq.~\eqref{spin_dependent+independent_tunneling},
the effective Josephson couplings between $d$ vectors $\mathbf{d}_{n}$ and $\mathbf{d}_m$ in Eq.~\eqref{josephson_couplings_decoupled_spin_orbital_expansion}
are given by
\begin{align}
    {J}_{nm} 
    & \approx  \nonumber
    - \frac{\Delta_0}{W^2}
    \underset{j \in \Sigma_m}{\sum_{i \in \Sigma_n,}}
    \bigg\{
    t_0^2  {-} 2 J_{sd}^2 \alpha_{ij} - 4i \frac{J_{sd}^3}{t_0} \chi_{ijk}
    \bigg \}
    + \mathcal{O}(J_{sd}^4),
    \\
    {\mathbf{D}}_{nm}
    &\approx \nonumber
    -2
    \frac{\Delta_0}{W^2}
    \underset{j \in \Sigma_m}{\sum_{i \in \Sigma_n,}}
    \bigg\{
    J_{sd}^2 \boldsymbol{\beta}_{ij}
    +
    2i \frac{J_{sd}^3}{t_0} \boldsymbol{\gamma}_{ijk}
    \bigg\}
    + \mathcal{O}(J_{sd}^4),
    \\
    {\Gamma}_{nm}^{ab}
    &\approx
    -2 
    \frac{\Delta_0}{W^2}
    \underset{j \in \Sigma_m}{\sum_{i \in \Sigma_n,}}
    \frac{J_{sd}^4}{t_0^2} \beta_{ij}^a \beta_{ij}^b
     + \mathcal{O}(J_{sd}^5),
     \label{triangular_lattice_josephson_coupling_amplitudes}
\end{align}
up to an overall factor.
Above, we approximate the weighting function
as $w_{nm} \sim - \Delta_0/W^2$,
in which $\Delta_0$ is the magnitude of the pairing gap function and $W$ is the bandwidth.

In the limit of vanishing coupling to the exchange field, the Heisenberg-like $J_{nm}$ term dominates, promoting collinear $d$ vectors at adjacent superconducting grains.
For $J_{nm}<0$, minimization of the free energy leads to a homogeneous order parameter, with any spatial variation of the $d$ vector being penalized.
For finite $J_{sd}$, it follows that  DM-like coupling $\mathbf{D}_{nm}$ can play a significant role and compete with $J_{nm}$.
This competition promotes noncollinear configurations of $d$ vectors at adjacent grains and can stabilize spatially inhomogeneous $d$ vector textures, such as skyrmion-like configurations.
The $\Gamma$-type coupling represents a higher order tunneling process and can further reinforce noncollinear $d$ vector configurations.

For frustrated magnetic textures, the dominant contributions stem from the Heisenberg-like and DM-like Josephson couplings, leading to noncollinear $d$ vectors at adjacent superconducting grains.
The relative angle between $d$ vectors is given by
\begin{equation}
\theta = \arctan (D_\perp/J_{nm}),
\end{equation}
in which
\begin{math}
    D_{\perp} = \mathbf{D}_{nm} \cdot {(\hat{\mathbf{d}}_n \times \hat{\mathbf{d}}_m^*)}/{|\hat{\mathbf{d}}_n \times \hat{\mathbf{d}}_m^*|}
\end{math}
is the component of $\mathbf{D}_{nm}$ normal to the plane spanned by $\mathbf{d}_n$ and $\mathbf{d}_m^*$.
In Fig.~\ref{fig:kspace_dvector_texture}, we show the nonuniform $d$ vector texture in a hybrid picture of momentum space describing the $d$ vectors for the superconducting grains, and the real-space Josephson coupling.
Finite DM-like Josephson coupling favors a spatially inhomogeneous pairing order, which can lead to nontrivial contributions to the superfluid velocity~\cite{Frazier2025}.

Both the relative strength of the $s$-$d$ exchange and the noncollinearity of the underlying spin texture can determine the characteristic length scale of the $d$ vector textures.
Namely, the ratio of DM-like  to Heisenberg-like Josephson couplings is given by
\begin{math}
    {\mathbf{D}_{nm}}/{J_{nm}} \propto 
    \sum_{i \in \Sigma_n, j \in \Sigma_m}
    J_{sd}^2
    \boldsymbol{\beta}_{ij},
\end{math}
to leading order in $J_{sd}$.
In, for example, 
$4H_b$-TaS$_2$, the $s$-$d$ exchange has been shown to be on the order of 10 meV~\cite{Vano2021}.
Below, we analyze three representative cases illustrating how different underlying spin configurations affect the Josephson coupling in the limit of $|J_{sd}| \ll W$.
When the $s$-$d$ coupling is comparable to the bandwidth ($|J_{sd}| \gtrsim |t_0|$), the perturbative expansion of the tunneling in Eq.~\eqref{effective_tunneling_Jsd_expansion} is not valid; rather, it is necessary to employ a new form of the Josephson couplings in Eq.~\eqref{josephson_couplings_decoupled_spin_orbital_expansion}, as discussed in Appendix~\ref{appendix:strong_Jsd_tunneling}.

\subsection{Collinear spin configuration} 

We first consider a system in which the local spins are in a ferromagnetic or collinear antiferromagnetic configuration.
In this case, the magnitude of $\alpha_{ij} = \mathbf{s}_i \cdot \mathbf{s}_j$ is maximized, while the measure of noncollinearity $\boldsymbol{\beta}_{ij} = \mathbf{s}_i \times \mathbf{s}_j$ and spin chirality $\chi_{ijk} = \mathbf{s}_i \cdot (\mathbf{s}_j \times \mathbf{s}_k)$ vanish.
The higher order term, $\boldsymbol{\gamma}_{ijk}$, is finite for collinear spin configuration, but contributes weakly to the Josephson free energy.
The vector $\mathbf{D}_{nm}$ does not vary in space and is uniformly aligned in the local direction of the N\'eel vector.
However, because the DM-like term originates from a higher order tunneling process, its effect is weak.
Consequently, for collinear spin configurations with weak $s$-$d$ exchange, the Heisenberg-like Josephson coupling $J_{nm}$ dominates, promoting a collinear $d$ vector texture.
Depending on the sign of $J_{nm}$, this will either favor a homogeneous superconducting order ($J_{nm}<0$) or a spatially modulating $\mathrm{U}(1)$ phase ($J_{nm}>0$).

\subsection{Coplanar spin configuration}

Next, we turn to the classical $120^\circ$ coplanar ordering of the three-sublattice antiferromagnet.
Here, the scalar spin chirality vanishes, $\chi_{ijk} = 0$, whereas $\alpha_{ij}$ and $\beta_{ij}$ are both finite.
To lowest order in $J_{sd}$, the nonvanishing terms in Josephson free energy in Eq.~\eqref{triangular_lattice_josephson_coupling_amplitudes} are given by
\begin{equation}
    \begin{aligned}
        {J}_{nm} 
        &\approx 
        -\frac{\Delta_0}{W^2}
        \underset{j \in \Sigma_m}{\sum_{i \in \Sigma_n,}}
        (
        t_0^2 {-} J_{sd}^2
        )
        \\
        {\mathbf{D}}_{nm} 
        &\approx
        {-} \frac{\Delta_0}{W^2}
        \underset{j \in \Sigma_m}{\sum_{i \in \Sigma_n,}}
        (\sqrt{3} J_{sd}^2 \nu
        \epsilon_{ij}
        \hat{z}
        ),
    \end{aligned}
\end{equation}
where  $\alpha_{ij} = \cos 2\pi/3 = -1/2$ and $\boldsymbol{\beta}_{ij} = (\sqrt{3}/2) \epsilon_{ij} \hat{z}$ for the $120^{\circ}$ ordered spin configuration, with $\epsilon_{ij}$ being the antisymmetric tensor.
The discrete superfluid stiffness $J_{nm}$ competes with the DM-like Josephson coupling $D_{nm}$, favoring noncollinear $d$ vectors at adjacent grains and leading to a spatially inhomogeneous superconducting pairing order.
For example, when $|J_{sd}/t_0| \sim 1/10$, it follows that the relative magnitudes of Josephson couplings are of order $|\mathbf{D}_{nm}|/J_{nm} \sim (J_{sd}/t_0)^2 \sim 10^{-2}.$
Consequently, the relative angle between $d$ vectors at adjacent grains is of order $\theta \sim 10^{-2}$.
The characteristic length of the $d$ vector textures is $\lambda = (2\pi/\theta)\xi$, in which $\xi$ is the size of the superconducting grain.
The superconducting grain size is generally set by the superconducting coherence length, which is on the order of tens of nanometers~\cite{Ribak2020, Jeon2021}.
The resulting $d$ vector texture has a characteristic length scale on the order of tens of microns.
Hence, even in the absence of scalar spin chirality, noncollinear spin configurations can promote frustrated $d$ vector textures.

\subsection{Noncoplanar spin configuration} 

Lastly, we consider the case where the three $120^\circ$ ordered spins in the ground state are canted out-of-plane by angle $\theta_\mathrm{0}$ in Eq.~\eqref{three_spins}.
Finite $\theta_0$ leads to nonvanishing scalar spin chirality,
\begin{equation}
    \chi_{abc} 
    =
    \nu \frac{3 \sqrt{3}}{16} \cos^2 \theta_0 \sin \theta_0,
\end{equation}
which is dependent on the sign of the canting angle and the vector spin chirality, $\nu$.
The corresponding spin-dependent factor $\boldsymbol{\beta}_{ij}$ acquires an in-plane component and is dependent on the intersublattice tunneling at the grain boundary, with
    \begin{align}
        \boldsymbol{\beta}_{ab} 
        &= \nonumber
        \frac{1}{8}
        \left(
        \begin{array}{c}
             \sin(2\theta_0) \Big( \frac{3}{2} \sin \varphi_0 -  \frac{\sqrt{3}}{2} \nu \cos \varphi_0 \Big)
             \\
             - \sin(2\theta_0) \Big(\frac{3}{2} \cos \varphi_0 +  \frac{\sqrt{3}}{2} \nu \sin \varphi_0 \Big)
             \\
             \nu \sqrt{3} \cos^2 \theta_0
        \end{array}
        \right),
        \\
        \boldsymbol{\beta}_{bc} 
        &=  \nonumber
        \frac{\nu \sqrt{3}}{8}
        \left(
        \begin{array}{c}
             \sin(2\theta_0) \cos \varphi_0
             \\
             \sin(2\theta_0) \sin \varphi_0
             \\
             \cos^2 \theta_0
        \end{array}
        \right),
        \\
        \boldsymbol{\beta}_{ca} 
        &= 
        \frac{1}{8}
        \left(
        \begin{array}{c}
             - \sin(2\theta_0) \Big( \frac{3}{2} \sin \varphi_0 +  \frac{\sqrt{3}}{2} \nu \cos \varphi_0 \Big)
             \\
             \sin(2\theta_0) \Big( 
             \frac{3}{2} \cos \varphi_0 - \frac{\sqrt{3}}{2} \nu \sin \varphi_0
             \Big)
             \\
             \nu \sqrt{3} \cos^2 \theta_0
        \end{array}
        \right).
    \end{align}
As a result, $\mathbf{D}_{nm}$ in Eq.~\eqref{triangular_lattice_josephson_coupling_amplitudes} changes sign and direction in real-space and is sensitive to the geometry of the superconducting grains.
Nonetheless, minimization of the Josephson free energy requires the $d$ vector to develop a spatially inhomogeneous texture, with characteristic length scale on the order of tens of microns, similar to when there is a coplanar spin configuration.

\section{Competition between bulk pairing order and Josephson coupling}

\label{sec:competition_of_bulk_and_josephson_coupling}

We now comment on the relevant energy scales in the free energy in Eq.~\eqref{free_energy_general}.
As discussed in Sec.~\ref{sec:josephson_coupling}, the total energy can be separated into two contributions: the condensation energy $F_\mathrm{homogeneous}$ describing the bulk pairing order for a single isolated superconducting grain and the Josephson coupling energy $F_\mathrm{variation}$ describing a spatially varying pairing order, corresponding to many coupled superconducting grains.

The term $F_\mathrm{homogeneous}$ describes the energy contribution associated with forming a uniform superconducting order parameter for a single superconducting grain and is associated with the condensation energy.
For a two-dimensional system, to lowest order, the condensation energy scales as $D(0) \Delta_0^2$, where $D(0)$ is the density of states at the Fermi level and $\Delta_0$ is the magnitude of the superconducting gap.
The term scales extensively with the area of the superconducting grain, and the lower bound for the grain size is generally set by the superconducting coherence length, $\xi$.
In contrast, $F_\mathrm{variation}$ describes the Josephson coupling across grain boundaries and scales extensively with the perimeter of the superconducting grain.
Hence, the size of the superconducting grains and the relative amplitude of Josephson tunneling play a key role in determining the real-space texture of the pairing order.

For large grains, with characteristic size $R \gg \xi$, $F_\mathrm{homogeneous}$ dominates, while the $F_\mathrm{variation}$ is relatively small.
In this case, the pairing order for the bulk is determined from the superconducting gap function and band structure, as discussed in Sec.~\ref{sec:pairing_correlations}, and the order parameter remains essentially uniform.
However, for small grains, in which $R$ is comparable to $\xi$, $F_\mathrm{variation}$ becomes comparable or exceeds $F_\mathrm{homogeneous}$.
This can lead to a spatially varying pairing order parameter, with the $d$ vector texture largely determined by the free energy in Eq.~\eqref{free_energy_josephson_coupling_discrete}.
In this regime, the $d$ vector orientation constitutes an additional degree of freedom and can give rise to, for example, contributions to the superfluid velocity or anomalous vortices~\cite{Frazier2025}.

Moreover, the relative strength of the different types of Josephson couplings is crucial.
The Heisenberg-like coupling, which favors collinear $d$ vector textures, scales as $\Delta_0t_0^2/W^2$ to lowest order and is negative-valued, as shown in Eq.~\eqref{triangular_lattice_josephson_coupling_amplitudes}.
For ${J_{nm}<0}$, this term does not compete with but rather reinforces homogeneous bulk pairing order.
By contrast,
the effective DM-like Josephson coupling $\mathbf{D}_{nm}$ arises from the $s$-$d$ exchange and scales with $\Delta_0 J_{sd}^2/W^2$ to lowest order.
For systems with very weak $s$-$d$ coupling, the DM-like term is negligible, and any spatial inhomogeneity is penalized.
The system will behave as a uniform bulk superconductor determined by $F_\mathrm{homogeneous}$.
However, for finite $J_{sd}$, the DM-like Josephson coupling competes with both the Heisenberg-like coupling and bulk condensation energy, promoting a spatially varying $d$ vector texture.

\section{Josephson diode effect arising from frustrated spin textures}

\label{sec:josephson_diode}

\begin{figure}
    \centering
    \includegraphics[width= \linewidth]{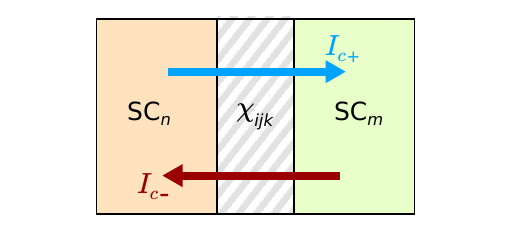}
    \caption{
    Spin chirality induced diode effect in an SIS junction between superconductors SC$_n$ (left) and SC$_m$ (right).
    Here, the barrier region (gray, striped) has a frustrated magnetic texture.
    Finite spin chirality in the barrier region, $\chi_{ijk}\neq 0$,  leads to direction-dependent critical currents $I_{c+}$ and $I_{c-}$, resulting in a Josephson diode effect.}
    \label{fig:josephson_diode}
\end{figure}

Lastly, we demonstrate that the presence of finite spin chirality in the underlying local exchange field or antisymmetric coupling of noncollinear $d$ vectors can lead to a Josephson diode effect.
Consider a superconductor-insulator-superconductor (SIS) Josephson junction between superconductors $n$ and $m$, as shown schematically in Fig.~\ref{fig:josephson_diode}, in which the barrier region hosts a frustrated spin texture.
Considering up to second order Josephson tunneling, the Josephson current can be expressed as
\begin{equation}
    I_J^{(n,m)}(\phi) = I_1 \sin(\phi-\phi_1) + I_2\sin(2\phi-\phi_2)
\end{equation}
in which $\phi = \phi_n - \phi_m$ is the difference in $\mathrm{U}(1)$ phases, $I_1$ and $I_2$ are the magnitudes of the first and second order Josephson currents, respectively, and $\phi_1$ and $\phi_2$ are constant phase shifts.
We take the approximation that the Josephson current is primarily dominated by the first order contribution.
As such, the Josephson critical current reaches extrema at
\begin{math}
    \phi_+ 
    \approx \nonumber 
    {\pi}/{2}+\phi_1,
\end{math}
and
\begin{math}
    \phi_- 
    \approx
    {3 \pi}/{2}+\phi_1
\end{math}
yielding critical current
\begin{math}
    I_{c+} 
    \approx 
    I_1 
    - I_2 \sin(
    2\phi_1 - \phi_2)
\end{math}
and
\begin{math}
    I_{c-} 
    \approx 
    -I_1 
    - I_2 \sin(
    2\phi_1 - \phi_2)
\end{math}
respectively.
It follows that the Josephson diode efficiency is given by
\begin{equation}
    \eta
    =
    \frac{|I_{c+}| - |I_{c-}|}{|I_{c+}| + |I_{c-}|}
    \approx
    \frac{I_2}{{I_1}} |\sin(2\phi_1 - \phi_2)|
    +
    \mathcal{O}\Big( \frac{I_2^2}{I_1^2} \Big).
    \label{josephson_diode_efficiency_general}
\end{equation}
Below, we demonstrate that when there is nonvanishing spin chirality or noncollinear $d$ vector configuration, this can lead to a Josephson diode effect.

From the Ambegaokar-Baratoff formalism, the first order Josephson current between grains $n$ and $m$ is given by~\cite{Ambegaokar1963, Mahan2000, Sigrist1991, Frazier2024}
\begin{equation}
    I_1(\phi)
    =
    i \frac{e}{\hbar} \left( \mathcal{J}_{nm} - \mathrm{c.c.}\right),
\end{equation}
with $\mathcal{J}_{nm}$ being the Josephson form factor in Eq.~\eqref{josephson_form_factor}.
For simplicity, we consider a Josephson junction between two unitary spin triplet superconductors, with pairing order of the $n\mathrm{th}$ grain described by $e^{i\phi_n} \hat{\mathbf{d}}_n$, in which $\phi_n$ is the $\mathrm{U}(1)$ superconducting phase and $\hat{\mathbf{d}}_n$ is the real dimensionless $d$ vector.

In addition to the difference in $\mathrm{U}(1)$ phases, the Josephson form factor is given by the relative orientation of $d$ vectors, as shown in Eq.~\eqref{josephson_form_factor}.
The first order Josephson current can be expressed as
\begin{equation}
    I_1(\phi) = I_1 \sin (\phi - \phi_1),
\end{equation}
in which
\begin{math}
    I_1 = (e/\hbar)|\tilde{\mathcal{J}}_{nm}|
\end{math}
and
\begin{math}
    \tan \phi_1 = -\mathrm{Im}(\tilde{\mathcal{J}}_{nm})/\mathrm{Re}(\tilde{\mathcal{J}}_{nm}).
\end{math}
Here, we define
\begin{equation}
    \tilde{\mathcal{J}}_{nm} = 
    J_{nm}
    \hat{\mathbf{d}}_n \cdot \hat{\mathbf{d}}_m
    +
    \mathbf{D}_{nm}
    \cdot
    (
    \hat{\mathbf{d}}_n \times \hat{\mathbf{d}}_m)
    +
    \hat{\mathbf{d}}_n^\mathrm{T} \Gamma_{nm} \hat{\mathbf{d}}_m,
    \label{tilde_J.phase_independent}
\end{equation}
which is independent of the $\mathrm{U}(1)$ phase difference and encodes the relative alignment of $d$ vectors.
From the Josephson couplings in Eq.~\eqref{triangular_lattice_josephson_coupling_amplitudes}, it follows that
\begin{align}
    \mathrm{Re}(\tilde{\mathcal{J}}_{nm})
    \approx \nonumber
    - \frac{\Delta_0}{W^2}
    &
    \sum_{i \in \Sigma_n, j \in \Sigma_m}
    \bigg\{
    (t_0^2  {-} 2 J_{sd}^2 \alpha_{ij}) \hat{\mathbf{d}}_n \cdot \hat{\mathbf{d}}_m
    \\
    \nonumber
    &+
    2 J_{sd}^2 \boldsymbol{\beta}_{ij}
    \cdot
    (\hat{\mathbf{d}}_n \times \hat{\mathbf{d}}_m)
    +
    \mathcal{O}(J_{sd}^4)
    \bigg \}
    \\
    \mathrm{Im}(\tilde{\mathcal{J}}_{nm})
    \approx \nonumber
    - \frac{4 \Delta_0}{W^2}
    &
    \sum_{i \in \Sigma_n, j \in \Sigma_m}
    \bigg\{
    - \frac{J_{sd}^3}{t_0} \chi_{ijk}
    (\hat{\mathbf{d}}_n \cdot \hat{\mathbf{d}}_m),
    \\
    &+
    \frac{J_{sd}^3}{t_0} \boldsymbol{\gamma}_{ijk} \cdot 
    (\hat{\mathbf{d}}_n \times \hat{\mathbf{d}}_m)
    +
    \mathcal{O}(J_{sd}^4)
    \bigg \}.
\end{align}
Here, we have taken the leading order terms in the $s$-$d$ coupling, which comprise the Heisenberg-like and DM-like Josephson couplings.
The imaginary component, cubic in $J_{sd}$ and corresponding to second order tunneling processes, produces a finite phase shift, resulting in a $\phi_0$ junction with $I_1(\phi)\neq -I_1(-\phi)$.

The second order Josephson effect corresponds to a term that is fourth order in the single-particle tunneling described by tunneling matrix $T_{nm}$.
Under the assumption that the spin-independent nearest-neighbour tunneling $t_0$ is most heavily weighted in the tunneling process, it follows that the second order Josephson critical current is given by
\begin{math}
    I_2(\phi) \approx I_2 \sin(2\phi),
\end{math}
in which
\begin{math}
    I_2
    \approx
    (e/\hbar) (N_{nm} \Delta_0 {t_0^4}/{W^4}) (\hat{\mathbf{d}}_n \cdot \hat{\mathbf{d}}_m)^2
\end{math}
to lowest order, with $N_{nm}$ being the dimensionless length of the junction interface between superconducting grains $n$ and $m$.
Here, the factor $t_0^4$ reflects the fourth-order tunneling process, and for simplicity, we have considered $\phi_2 \approx 0$.

It follows that the Josephson diode efficiency in Eq.~\eqref{josephson_diode_efficiency_general} is given by
\begin{equation}
    \eta \approx 
    \frac{N_{nm} \Delta_0 t_0^4 (\hat{\mathbf{d}}_n \cdot \hat{\mathbf{d}}_m)^2} {W^4 |\tilde{\mathcal{J}}_{nm}|}
    |\sin 2\phi_1|,
\end{equation}
in which
\begin{equation}
    \phi_1 = 
    - \arctan 
    \left(\frac{\mathrm{Im} (\tilde{\mathcal{J}}_{nm})}{\mathrm{Re}(\tilde{\mathcal{J}}_{nm})}
    \right)
    \approx
    - \frac{\mathrm{Im} (\tilde{\mathcal{J}}_{nm})}{\mathrm{Re}(\tilde{\mathcal{J}}_{nm})}.
\end{equation}
For the three-sublattice system, the diode efficiency reduces to
\begin{align}
    \eta
    &\approx \nonumber
    \frac{2
    N_{nm}
    \Delta_0 t_0^4 (\hat{\mathbf{d}}_n \cdot \hat{\mathbf{d}}_m)^2}{W^4 |\tilde{\mathcal{J}}_{nm}|}
    \left|
    \frac{\mathrm{Im} (\tilde{\mathcal{J}}_{nm})}{\mathrm{Re}(\tilde{\mathcal{J}}_{nm})}
    \right|
    \\
    &\propto
    \bigg|
    \underset{j \in \Sigma_m}{\sum_{i \in \Sigma_n,}}
    \bigg\{
    \chi_{ijk}
    (\hat{\mathbf{d}}_n \cdot \hat{\mathbf{d}}_m)
    -
    \boldsymbol{\gamma}_{ijk} \cdot 
    (\hat{\mathbf{d}}_n \times \hat{\mathbf{d}}_m)
    \bigg \}
    \bigg|.
    \label{diode_efficiency_three_sublattice}
\end{align}
For collinear $d$ vectors in grains $n$ and $m$, the diode efficiency scales linearly with the spin chirality in the barrier region.
For noncollinear $d$ vectors, the diode efficiency is additionally dependent on the time-reversal breaking factor $\boldsymbol{\gamma}_{ijk}$ in Eq.~\eqref{spin_factors} and relative orientation of $d$ vectors.

The Josephson diode effect arises from the effective tunneling in the presence of a frustrated local spin texture locally breaking time reversal and inversion symmetries, which are both necessary to produce a finite diode effect~\cite{Wu2022, Davydova2022, Zhang2022a, Wang2025}.
Here, the scalar spin chirality is antisymmetric, with $\hat{\mathbf{s}}_i\cdot (\hat{\mathbf{s}}_j \times \hat{\mathbf{s}}_k) = - \hat{\mathbf{s}}_k\cdot (\hat{\mathbf{s}}_j \times \hat{\mathbf{s}}_i)$, corresponding to local inversion symmetry breaking for noncoplanar spin configurations.
However, unlike the spin factor $\boldsymbol{\beta}_{ij}$ in Eq.~\eqref{spin_factors}, the spin chirality is a scalar term and hence enters into the spin-independent tunneling of electrons, as shown in Eq.~\eqref{spin_dependent+independent_tunneling}.
As such, the tunneling of itinerant electrons in the presence of nonvanishing spin chirality breaks time reversal symmetry in addition to local inversion symmetry.
For the latter term in Eq.~\eqref{diode_efficiency_three_sublattice}, $\boldsymbol{\gamma}_{ijk}$ is odd under time-reversal but invariant under inversion, $\gamma_{ijk} = \gamma_{jik}$.
However, it is coupled to $\hat{\mathbf{d}}_n \times \hat{\mathbf{d}}_m$, which, for noncollinear $d$ vector configuration, breaks inversion,  $\hat{\mathbf{d}}_n \times \hat{\mathbf{d}}_m = - \hat{\mathbf{d}}_m \times \hat{\mathbf{d}}_n$.
Hence, the Josephson diode effect can arise from noncoplanar spin structure in the barrier region, or from a noncollinear $d$ vector configuration.

Moreover, when the underlying spin configuration has nonvanishing spin chirality, this is also applicable to, for example, junctions between two spin singlet superconductors. 
For spin singlet superconductors, there is only a $\mathrm{U}(1)$ degree of freedom, and as such, the scalar Josephson coupling, in analogy to the Heisenberg-like coupling $J_{nm}$, will persist.
For the same symmetry reasons outlined above, the effective tunneling in the presence of a frustrated spin texture in the barrier region will lead to a Josephson diode effect with efficiency proportional to the scalar spin chirality of the underlying magnetic texture.

For example, consider the case of collinear $d$ vectors ($\hat{\mathbf{d}}_n \times \hat{\mathbf{d}}_m = 0$).
From Eq.~\eqref{diode_efficiency_three_sublattice}, the diode efficiency is given by
\begin{equation}
    \eta 
    \approx
    \frac{8 |t_0 J_{sd}|^3 \bar{\chi}}
    {
    W^2 
    (t_0^2 - 2J_{sd}^2 \bar{\alpha})
    \sqrt{(2J_{sd}^2 \bar{\alpha} -t_0^2)^2 + (J_{sd}^3 \bar{\chi}/t_0 )^2}},
\end{equation}
in which  $\bar{\chi}$ and $\bar{\alpha}$ denote the averages of $\chi_{ijk}$ and $\alpha_{ij}$, respectively, over the junction interface.
This expression simplifies to $\eta \approx 
8  \bar{\chi} |J_{sd}^3/(t_0 W^2)|$
to leading order.
For $|J_{sd}| \sim |t_0|/10$ and a strongly noncoplanar spin structure with $\bar{\chi} \sim \mathcal{O}(1)$, it follows that the diode efficiency is on the order of $10^{-3}.$
In the nonperturbative regime, when $J_{sd}$ is comparable to $t_0$, the Josephson diode effect persists and can yield a larger diode efficiency.
Considering higher order terms in $J_{sd}$ in the Josephson couplings in Eq.~\eqref{triangular_lattice_josephson_coupling_amplitudes}, there will be additional contributions to the critical current which are dependent on the underlying spin configuration.
Consequently, the imaginary and real parts of the Josephson form factor can be of comparable magnitude, which can impact both the first and second order critical currents.
We discuss the resulting diode efficiency in the limit of $|J_{sd}| \gg |t_0|$ in Appendix~\ref{appendix:strong_Jsd_tunneling}.
While the Josephson diode efficiency can differ in the regime of large $s$-$d$ coupling, the essential features remain: specifically, terms that break both time-reversal and inversion symmetry will give rise to a finite diode effect.

Lastly, we comment on the effects of Majorana edge states.
In systems that admit spin triplet pairing correlations, the pairing order transforms according to a nontrivial representation of orbital angular momentum.
For a system with only spin and orbital degrees of freedom, spin triplet pairing order
has odd integer partial wave symmetry in momentum space.
This can lead to Majorana states, which can contribute, for example, $4\pi$-periodic Josephson current when fermion parity is conserved at the junction barrier~\cite{Kitaev2001, Kwon2003, Fu2008, Chung2009, Sau2010, Alicea2010, Grosfeld2011, Alicea2012}.
In this case, the Josephson current can be expressed as
\begin{math}
    I_J^{(n,m)}(\phi) \approx I_0 \sin(\phi/2-\phi_0) + I_1 \sin(\phi-\phi_1) + I_2\sin(2\phi-\phi_2)
\end{math}
to lowest order.
The magnitude of $4\pi$-periodic Josephson current, $I_0$, can be obtained, for example, by projecting to the Majorana states localized at the junction barrier.
The contribution from the $4\pi$-periodic Josephson current can modify the critical currents $I_{c+}$ and $I_{c-}$ and is sensitive to the relative magnitude of $I_0$ and $I_1$.
Nonetheless, the discrepancy in $I_{c+}$ and $I_{c-}$ arises from the admixture of different-order Josephson currents.
The salient features of the Josephson diode effect, which arise in the presence of broken parity and time reversal symmetry at the junction interface, persist.

\section{Conclusions and Discussion}

We have demonstrated that coupling itinerant electrons in a spin triplet superconductor to a local exchange field consisting of frustrated spins can generate anisotropic Josephson couplings between $d$ vectors.
These anisotropic Josephson couplings, analogous to Dzyaloshinskii-Moriya and $\Gamma$-type interactions in magnetism, endow a ``pliability'' to the pairing order that competes with the superfluid stiffness and can stabilize a spatially varying $d$ vector texture.

Such terms correspond microscopically to Josephson coupling of superconducting grains and can be realized when itinerant electrons in a spin triplet superconductor are coupled to a frustrated spin texture.
Local spins in an $s$-$d$ model on a geometrically frustrated lattice lead to an effective tunneling of itinerant electrons that is dependent on the underlying magnetic configuration.
Moreover, spin triplet pairing correlations can arise either from $s$-$d$ exchange or spin-orbit coupling.
The presence of a noncollinear frustrated spin texture for the three-sublattice system leads to anisotropic DM-like and $\Gamma$-type Josephson couplings, which can promote spatially varying $d$ vector textures.
Lastly, a Josephson diode effect can arise when the junction barrier hosts noncoplanar spins with finite spin chirality, or when there is antisymmetric Josephson coupling between noncollinear $d$ vectors, breaking both time-reversal and inversion symmetry at the interface between superconducting grains.

The results are relevant to recent experiments suggesting the coexistence of frustrated magnetism and unconventional superconductivity.
For example, in kagome noncollinear antiferromagnet Mn$_3$Ge, where unconventional superconductivity emerges in proximity to Nb, spins of Mn serve as a local exchange field that can mediate DM-like and $\Gamma$-type Josephson couplings, even for coplanar spin configurations.
For SIS junctions consisting of Mn$_3 X$ ($X$ = Sn, Ge) in the junction barrier region between two $s$-wave superconductors, this can lead to a Josephson diode effect for noncoplanar spin configurations.

Additionally, the results are pertinent to superconducting $4H_b$-TaS$_2$.
The frustrated spin textures in the $1T$-TaS$_2$ layers can serve as a local exchange field for the Ising superconducting $1H$-TaS$_2$ layers.
Here, Ising spin-orbit coupling, in addition to the coupling to frustrated magnetic textures, can generate spin triplet superconducting pairing correlations.
The competition of Ising spin-orbit coupling and the local spin textures can lead to spatially varying $d$ vector textures in absence of an external field.

The results of this work, demonstrating the existence of anisotropic Josephson couplings arising from interplay with frustrated spin configurations, open avenues to understanding the origin of nontrivial spatial textures in unconventional superconductors.
While the present work treats the local spins as a static classical exchange field, the effects of quantum fluctuations and dynamic spin textures, for example, remain an open question to be explored.
Moreover, other forms of spin-dependent tunneling between superconducting grains can serve as a basis for realizing the proposed anisotropic Josephson couplings.

\section*{Acknowledgments}
We thank Junyi Zhang for helpful discussions at the early stage of this work. 
We also thank Oleg Tchernyshyov for reading through our manuscript and for 
helpful comments about the residual degrees of freedom of the $d$ vector textures and the perturbative methods. 
We acknowledge the support of the NSF CAREER Grant No.~DMR-1848349.

\appendix
\section{Josephson form factor and free energy}

\label{appendix:Josephson_form_factor}

In this appendix, we provide the Josephson form factor which describes the first order Josephson coupling between two superconducting grains.
At zero voltage bias, the Josephson form factor is given by~\cite{Ambegaokar1963, Geshkenbein1986, Sigrist1991, Frazier2024}
\begin{align}
    \nonumber
    \mathcal{J}_{n m}=
    - \frac{1}{\beta}
    &\sum_{i\omega_n}
    \sum_{\mathbf{k}, \mathbf{k}'}
    \mathrm{Tr}
    \Big[ \mathcal{F}_{n}(\mathbf{k}; i\omega_n) [T_{n m}(-\mathbf{k}, -\mathbf{k}'; i\omega_n)]^\mathrm{T}
    \\
    &\times
    [\mathcal{F}_{m}^\dagger(\mathbf{k}'; i\omega_n)]^\mathrm{T} T_{n m}(\mathbf{k}, \mathbf{k}'; i\omega_n)
    \Big],
    \label{appendixEq:form_factor}
\end{align}
in which the trace is taken over internal degrees of freedom of the Cooper pair (\textit{e.g.} spin, sublattice, \textit{etc.}), and the summation is over fermionic Matsubara frequencies.
Here, $\mathcal{F}_n(\mathbf{k}; i\omega_n)$ is the anomalous Green's function of grain $n$, given in Eq.~\eqref{anomalous_GF.freq_space} in the main text.
Focusing on the spin degrees of freedom, the anomalous Green's function can be expanded into its spin singlet and spin triplet components as
\begin{equation}
    \mathcal{F}_{n}(\mathbf{k}; i\omega_n)
    =
    \Big( f_{n, 0} (\mathbf{k}; i\omega_n) + \mathbf{f}_n(\mathbf{k}; i\omega_n) \cdot \boldsymbol{\sigma} \Big)
    i\sigma^y.
\end{equation}
Substituting into Eq.~\eqref{appendixEq:form_factor}, the trace can be expressed as three distinct contributions,
\begin{equation}
    \mathrm{Tr}\Big[ \cdots \Big]
    =
    2
    \Big(
    C_{\mathrm{sing}-\mathrm{sing}}
    +
    C_{\mathrm{sing}-\mathrm{trip}}
    +
    C_{\mathrm{trip}-\mathrm{trip}}
    \Big),
\end{equation}
which correspond to the Josephson tunneling between spin singlet components, between spin triplet components, and between spin singlet and spin triplet components respectively.
The three contributions to the Josephson form factor are given by~\cite{Sigrist1991, Frazier2024}
\begin{widetext}
\begin{subequations}
    \begin{equation}
    \begin{aligned}
        C_{\mathrm{sing}-\mathrm{sing}}
        =
        &
        - f_{n, 0}(\mathbf{k}')
        f^*_{m, 0}(\mathbf{k}) 
        T_{n m; 0}(-\mathbf{k}, -\mathbf{k}')
        T_{n m; 0}(\mathbf{k}, \mathbf{k}')
        + f_{n, 0}(\mathbf{k}') f^*_{m, 0}(\mathbf{k})  \Big[ \mathbf{T}_{n m}(-\mathbf{k}, -\mathbf{k}') \cdot \mathbf{T}_{n m}(\mathbf{k}, \mathbf{k}')\Big],
    \end{aligned}
    \end{equation}
    \begin{equation}
    \begin{aligned}
        C_{\mathrm{sing}-\mathrm{trip}}
        =
        & \
        \Big[ \mathbf{f}_n(\mathbf{k}') \cdot \mathbf{T}_{n m}(-\mathbf{k}, -\mathbf{k}')\Big] f^*_{m, 0}(\mathbf{k}) T_{n m; 0}(\mathbf{k}, \mathbf{k}')
        +
        f_{n, 0}(\mathbf{k}') T_{n m; 0}(-\mathbf{k}, -\mathbf{k}') \Big[ \mathbf{f}_m^*(\mathbf{k}) \cdot \mathbf{T}_{n m}(\mathbf{k}, \mathbf{k}') \Big]
        \\
        &
        - T_{n m; 0}(-\mathbf{k}, -\mathbf{k}') f^*_{m, 0}(\mathbf{k}) \Big[ \mathbf{f}_n(\mathbf{k}') \cdot \mathbf{T}_{n m}(\mathbf{k}, \mathbf{k}')\Big]
        - 
        f_{n, 0}(\mathbf{k}') T_{n m; 0}(\mathbf{k}, \mathbf{k}')
        \Big[\mathbf{f}_m^*(\mathbf{k}) \cdot \mathbf{T}_{n m}(-\mathbf{k}, -\mathbf{k}')\Big]
        \\
        &+if_{n, 0}(\mathbf{k}') \mathbf{f}_m^*(\mathbf{k}) \cdot \Big[ \mathbf{T}_{n m}(-\mathbf{k}, -\mathbf{k}') \times \mathbf{T}_{n m}(\mathbf{k}, \mathbf{k}')\Big]
        + if^*_{m, 0}(\mathbf{k}) \mathbf{f}_n(\mathbf{k}') \cdot \Big[\mathbf{T}_{n m}(-\mathbf{k}, -\mathbf{k}') \times \mathbf{T}_{n m}(\mathbf{k}, \mathbf{k}')\Big],
    \end{aligned}
    \end{equation}
    \begin{equation}
    \begin{aligned}
        C_{\mathrm{trip}-\mathrm{trip}}
        = 
        & \
        T_{n m; 0}(-\mathbf{k}, -\mathbf{k}')  T_{n m; 0}(\mathbf{k}, \mathbf{k}') \Big[ \mathbf{f}_n(\mathbf{k}') \cdot \mathbf{f}_m^*(\mathbf{k})\Big]
        + 
        \Big[ \mathbf{T}_{n m}(-\mathbf{k}, -\mathbf{k}') \cdot \mathbf{T}_{n m}(\mathbf{k}, \mathbf{k}') \Big]
        \Big[ \mathbf{f}_n(\mathbf{k}') \cdot\mathbf{f}_m^*(\mathbf{k}) \Big]
        \\
        &
        + iT_{n m; 0}(\mathbf{k}, \mathbf{k}') \mathbf{T}_{n m}(-\mathbf{k}, -\mathbf{k}') \cdot \Big[ \mathbf{f}_n(\mathbf{k}') \times \mathbf{f}_m^*(\mathbf{k}) \Big]
        +
        iT_{n m; 0}(-\mathbf{k}, -\mathbf{k}')  \mathbf{T}_{n m}(\mathbf{k}, \mathbf{k}') \cdot \Big[ \mathbf{f}_n(\mathbf{k}') \times \mathbf{f}_m^*(\mathbf{k})  \Big]
        \\
        &
        -
        \Big[\mathbf{f}_n(\mathbf{k}') \cdot \mathbf{T}_{n m}(-\mathbf{k}, -\mathbf{k}')\Big]
        \Big[\mathbf{f}_m^*(\mathbf{k}) \cdot \mathbf{T}_{n m}(\mathbf{k}, \mathbf{k}')\Big]
        -
        \Big[\mathbf{f}_n(\mathbf{k}') \cdot \mathbf{T}_{n m}(\mathbf{k}, \mathbf{k}')\Big]
        \Big[\mathbf{T}_{n m}(-\mathbf{k}, -\mathbf{k}') \cdot \mathbf{f}_m^*(\mathbf{k})\Big].
    \end{aligned}
    \end{equation}
    \label{appendixEq:form_factor_expansion}
\end{subequations}
\end{widetext}

For systems that admit a mixture of spin singlet and spin triplet correlations from broken parity symmetry, both $C_{\mathrm{sing}-\mathrm{trip}}$ and $C_{\mathrm{trip}-\mathrm{trip}}$ can contribute to the $d$ vector textures.
When tunneling satisfies $T_{nm}(\mathbf{k}, \mathbf{k}') = T_{nm}(-\mathbf{k}, -\mathbf{k}')$, it follows that $C_{\mathrm{sing}-\mathrm{trip}}$ will vanish due to singlet and triplet parts transforming oppositely under parity symmetry, $\mathbf{f}_{n}(\mathbf{k}) = -\mathbf{f}_{n}(-\mathbf{k})$ and $f_{n,0}(\mathbf{k}) = f_{n,0}(-\mathbf{k})$.
For the case of $T_{nm}(\mathbf{k}, \mathbf{k}') \neq T_{nm}(-\mathbf{k}, -\mathbf{k}')$, $C_{\mathrm{sing}-\mathrm{trip}}$ favors antiparallel $d$ vectors parallel to $\hat{\mathbf{n}} = T_{n m; 0}(-\mathbf{k}, -\mathbf{k}') \mathbf{T}_{nm}(\mathbf{k}, \mathbf{k}') - T_{n m; 0}(\mathbf{k}, \mathbf{k}') \mathbf{T}_{nm}(-\mathbf{k}, -\mathbf{k}')$ and will contribute to the Heisenberg-like Josephson coupling between $d$ vectors.
In the present work, we focus on the effective Josephson coupling between spin triplet correlations and thus analyze the last contribution, $C_{\mathrm{trip}-\mathrm{trip}}$.

Nonetheless, the coupling of spin singlet and spin triplet correlations, which arise in the presence of effective spin-dependent tunneling of itinerant electrons, can further promote spatially inhomogeneous spin triplet pairing order.
Namely, the $d$ vector can couple to the spin-dependent tunneling $\mathbf{T}_{nm}(\mathbf{k}, \mathbf{k}')$, or to the vector product $\mathbf{T}_{nm}(\mathbf{k}, \mathbf{k}') \times \mathbf{T}_{nm}(-\mathbf{k}, -\mathbf{k}')$.
For example, consider the tunneling in the presence of a frustrated local exchange field, given in Eq.~\eqref{spin_dependent+independent_tunneling}.
The coupling to $\mathbf{T}_{nm}$ is sensitive to the tunneling at the junction interface but nonetheless can further stabilize spatially inhomogeneous $d$ vector textures.

\section{Magnitude of Josephson coupling}

\label{appendix:weighting_factor}

In this appendix, we provide an estimate of the magnitude of the Josephson coupling.
Consider the factor $w_{nm}(\beta)$ in Eq.~\eqref{weighting_factor.sum_over_k}, which determines the weighting of the Josephson couplings.
For simplicity, suppose that the band structure of the normal metal is the same fofr grains $n$ and $m$.
Replacing the sum over momenta $\mathbf{k}$ and $\mathbf{k}'$ to an integral over the energy of the normal states, it follows that
\begin{align}
    w_{nm}(\beta)
    =
    \iint \mathrm{d}\xi \mathrm{d}\xi'
    \, D(\xi) D(\xi')
    \left\langle
    v(\mathbf{k}, \mathbf{k}')
    \right \rangle_{\xi, \xi'},
\end{align}
in which $D(\xi)$ is the single-particle density of states, and at zero-temperature,
\begin{equation}
    v(\mathbf{k}, \mathbf{k}')
    =-\frac{\Delta_0^2}{2}\frac{g_n(\mathbf{k}) g^*_m(\mathbf{k}')
    h(\mathbf{k}, \mathbf{k}')h(-\mathbf{k}, -\mathbf{k}')}{E_{n, \mathbf{k}} E_{m, \mathbf{k}'} (E_{n, \mathbf{k}} + E_{m, \mathbf{k}'})}.
\end{equation}
Above,
$\langle v(\mathbf{k}, \mathbf{k}') \rangle_{\xi, \xi'}$ denotes the average of $v(\mathbf{k}, \mathbf{k}')$
over the constant-energy surfaces of $\xi$ and $\xi'$ for momenta $\mathbf{k}$ and $\mathbf{k}'$ respectively.

In the weak-coupling regime, states at the Fermi surface primarily contribute to Josephson tunneling, with $D(\xi) \approx D(0) \sim 1/W$, in which $W$ is the bandwidth.
Suppose that the BdG quasiparticle energy for grain $n$ is $E_{n, \mathbf{k}} = \sqrt{\xi^2(\mathbf{k}) + |\Delta_0 g_n(\mathbf{k})|^2}$.
It follows that
\begin{align}
    w_{nm}
    \approx
    -
    \frac{\Delta_0 D^2(0)}{2}
    I_{nm},
\end{align}
in which
\begin{widetext}
\begin{align}
    I_{nm} = 
    \iint \mathrm{d}x \, \mathrm{d}x' 
    \left\langle
    \frac{g_n(\mathbf{k}) g_m^*(\mathbf{k}') h(\mathbf{k}, \mathbf{k}')
    h(-\mathbf{k}, -\mathbf{k}')}{
    \sqrt{x^2 + |g_n(\mathbf{k})|^2}
    \sqrt{{x'}^2 + |g_m(\mathbf{k}')|^2}
    \left(
    \sqrt{x^2 + |g_n(\mathbf{k})|^2} 
    + \sqrt{{x'}^2 + |g_m(\mathbf{k}')|^2}
    \right)}
    \right \rangle_{x, x'}
\end{align}
\end{widetext}
with $x = \xi/\Delta_0$ and $x' = \xi'/\Delta_0$.
Given nonvanishing overlap of $g_{n}(\mathbf{k})$ and $g_m(\mathbf{k}')$, the integral $I_{mn}$ is of order unity, and as such, the weighting factor scales as
\begin{equation}
    w_{nm}
    \sim
    -D^2(0) \Delta_0 \sim -\frac{\Delta_0}{W^2}.
\end{equation}
Consequently, provided that the tunneling matrices $T_{ij}$ are comparable to the hopping amplitude, the Josephson coupling amplitudes in Eq.~\eqref{josephson_couplings_decoupled_spin_orbital_expansion} are on the order of the pairing amplitude $\Delta_0$.

\section{Effective tunneling in the limit of strong \texorpdfstring{$s$-$d$}{s-d} exchange}

\label{appendix:strong_Jsd_tunneling}

In this section, we describe the effective tunneling in the presence of $s$-$d$ exchange in the $|J_{sd}| \gg |t_0|$ limit.
Consider the $s$-$d$ model
\begin{align}
    H 
    &= \nonumber
    -\mu
    \sum_{i, \alpha} c^\dagger_{i, \alpha} c_{i, \alpha}
    +
    t_0 \sum_{\langle ij\rangle, \alpha} c^\dagger_{i, \alpha} c_{j, \alpha}
    \\
    &\hspace{1em}+
    J_{sd} \sum_{i, \alpha, \beta}
    c^\dagger_{i, \alpha} \left[ \mathbf{s}_i \cdot \boldsymbol{\sigma} \right]_{\alpha \beta} c_{i, \beta},
\end{align}
as detailed in Sec.~\ref{sec:models_and_tunneling}.
In the limit of strong exchange coupling $|J_{sd}| \gg |t_0|$, the electrons are polarized according to the local exchange field $\hat{\mathbf{s}}_i$.
The state at the $i\mathrm{th}$ site is represented by
\begin{equation}
    | \hat{\mathbf{n}}_i\rangle
    =
    \left(
    \begin{array}{c}
         \cos \frac{\theta_i}{2}
         \\
         e^{i \varphi_i} \sin \frac{\theta_i}{2}
    \end{array}
    \right)
    \label{strong_J.eigenstates}
\end{equation}
up to a $\mathrm{U}(1)$ phase.
Here, $\hat{\mathbf{n}}_i = \langle \hat{\mathbf{n}}_i | \boldsymbol{\sigma} | \hat{\mathbf{n}}_i \rangle = 
( \sin \theta_i \cos \varphi_i, \sin \theta_i \sin \varphi_i, \cos \theta_i )^\mathrm{T}$
is the local magnetization of the itinerant electron at site $i$.
For antiferromagnetic coupling ($J_{sd} > 0$), $\hat{\mathbf{n}}_i = -\hat{\mathbf{s}}_i$, whereas for ferromagnetic coupling ($J_{sd} < 0$), $\hat{\mathbf{n}}_i = \hat{\mathbf{s}}_i$.
Treating the spin-independent nearest neighbour hopping perturbatively, the effective hopping matrix element between neighbouring sites $i$ and $j$ as~\cite{Ye1999, Ohgushi2000}
\begin{equation}
    t_{ij}^\mathrm{eff} = t_0 \langle \hat{\mathbf{n}}_i | \hat{\mathbf{n}}_j\rangle
    =
    t_0 |\langle \hat{\mathbf{n}}_i | \hat{\mathbf{n}}_j\rangle|
    e^{i \mathrm{arg}(\langle \hat{\mathbf{n}}_i | \hat{\mathbf{n}}_j\rangle)}
    \label{appendixEq:effective_t_ij.band_basis}
\end{equation}
in which
\begin{equation}
    |\langle \hat{\mathbf{n}}_i | \hat{\mathbf{n}}_j\rangle|
    =
    \sqrt{\frac{1 + \hat{\mathbf{n}}_i \cdot \hat{\mathbf{n}}_j}{2}}.
\end{equation}
The complex phase arises from the geometric gauge field, with
\begin{math}
    \arg (\langle \hat{\mathbf{n}}_i | \hat{\mathbf{n}}_j\rangle)
    \sim
    \int_i^j \mathbf{a}(\boldsymbol{\ell}) \cdot \mathrm{d}\boldsymbol{\ell},
\end{math}
with $\mathbf{a}(\boldsymbol{\ell})$ being the vector potential, and can lead to an anomalous quantum Hall effect for noncoplanar spin configurations~\cite{Nagaosa2010}.

Next, we write the effective tunneling in the $|J_{sd}| \gg |t_0|$ limit in the spin-basis.
In the band-diagonal basis, the hopping matrix element is given by
\begin{equation}
    T_{ij}^{(b)} 
    =
    t_0
    \left(
    \begin{array}{cc}
         \langle \hat{\mathbf{n}}_i|\hat{\mathbf{n}}_j\rangle
         &
         \langle \hat{\mathbf{n}}_i|-\hat{\mathbf{n}}_j\rangle
         \\
         \langle -\hat{\mathbf{n}}_i|\hat{\mathbf{n}}_j\rangle
         &
         \langle -\hat{\mathbf{n}}_i|-\hat{\mathbf{n}}_j\rangle
    \end{array}
    \right).
\end{equation}
The projected hopping matrix between nearest neighbours $i$ and $j$ in the band-diagonal basis is given by
\begin{align}
    PT_{ij}^{(b)}
    =
    \left(
    \begin{array}{cc}
         t^\mathrm{eff}_{ij}
         & 0  
         \\
         0 & 0
    \end{array}
    \right),
\end{align}
with $t_{ij}^\mathrm{eff}$ given in Eq.~\eqref{appendixEq:effective_t_ij.band_basis}.
As such, the projected hopping matrix, in the spin-basis, is given by
\begin{equation}
    T_{ij}^\mathrm{eff}
    =
    t_0 \langle \hat{\mathbf{n}}_i| \hat{\mathbf{n}}_j\rangle
    \left(
    \begin{array}{cc}
        \langle 
        \uparrow \!
        |
        \hat{\mathbf{n}}_i
        \rangle
        \langle \hat{\mathbf{n}}_j
        | \!
        \uparrow
        \rangle
        &
        \langle 
        \uparrow \!
        |
        \hat{\mathbf{n}}_i
        \rangle
        \langle \hat{\mathbf{n}}_j
        | \!
        \downarrow
        \rangle
        \\
        \langle 
        \downarrow \!
        |
        \hat{\mathbf{n}}_i
        \rangle
        \langle \hat{\mathbf{n}}_j
        | \!
        \uparrow
        \rangle
        &
        \langle 
        \downarrow \!
        |
        \hat{\mathbf{n}}_i
        \rangle
        \langle \hat{\mathbf{n}}_j
        | \!
        \downarrow
        \rangle
    \end{array}
    \right).
\end{equation}
This can be decomposed into Pauli matrices $T_{ij}^\mathrm{eff} = T_{ij; 0}^\mathrm{eff} + \mathbf{T}_{ij}^\mathrm{eff}\cdot\boldsymbol{\sigma}$, in which
\begin{equation}
    \begin{aligned}
        T_{ij; 0}^\mathrm{eff}
        &= 
        \frac{t_0}{2}
        \langle \hat{\mathbf{n}}_i| \hat{\mathbf{n}}_j\rangle
        \Big(
        \langle \hat{\mathbf{n}}_j
        | \!
        \uparrow
        \rangle
        \langle 
        \uparrow \!
        |
        \hat{\mathbf{n}}_i
        \rangle
        +
        \langle \hat{\mathbf{n}}_j
        | \!
        \downarrow
        \rangle
        \langle 
        \downarrow \!
        |
        \hat{\mathbf{n}}_i
        \rangle
        \Big),
        \\
        T_{ij;x}^\mathrm{eff}
        &=
        \frac{t_0}{2}
        \langle \hat{\mathbf{n}}_i| \hat{\mathbf{n}}_j\rangle
        \Big(
        \langle \hat{\mathbf{n}}_j
        | \!
        \downarrow
        \rangle
        \langle 
        \uparrow \!
        |
        \hat{\mathbf{n}}_i
        \rangle
        +
        \langle \hat{\mathbf{n}}_j
        | \!
        \uparrow
        \rangle
        \langle 
        \downarrow \!
        |
        \hat{\mathbf{n}}_i
        \rangle
        \Big),
        \\
        T_{ij;y}^\mathrm{eff}
        &=
        \frac{it_0}{2}
        \langle \hat{\mathbf{n}}_i| \hat{\mathbf{n}}_j\rangle
        \Big(
        \langle \hat{\mathbf{n}}_j
        | \!
        \downarrow
        \rangle
        \langle 
        \uparrow \!
        |
        \hat{\mathbf{n}}_i
        \rangle
        -
        \langle \hat{\mathbf{n}}_j
        | \!
        \uparrow
        \rangle
        \langle 
        \downarrow \!
        |
        \hat{\mathbf{n}}_i
        \rangle
        \Big),
        \\
        T_{ij;z}^\mathrm{eff}
        &=
        \frac{t_0}{2}
        \langle \hat{\mathbf{n}}_i| \hat{\mathbf{n}}_j\rangle
        \Big(
        \langle \hat{\mathbf{n}}_j
         | \!
         \uparrow
         \rangle
         \langle 
         \uparrow \!
         |
         \hat{\mathbf{n}}_i
         \rangle
         -
         \langle \hat{\mathbf{n}}_j
         | \!
         \downarrow
         \rangle
         \langle 
         \downarrow \!
         |
         \hat{\mathbf{n}}_i
         \rangle
        \Big).
    \end{aligned}
    \label{appendixEq:effective_tunneling.stong_Jsd_limit.decomposition}
\end{equation}
Considering the summation of tunneling processes at the boundary between two superconducting grains, the effective tunneling matrices in Eq.~\eqref{appendixEq:effective_tunneling.stong_Jsd_limit.decomposition} correspond to the tunneling matrices in Eq.~\eqref{josephson_form_factor}.

Notably, a frustrated magnetic texture is not necessary to achieve the anisotropic Josephson couplings in Eq.~\eqref{free_energy_josephson_coupling_discrete} in the limit of strong $s$-$d$ coupling.
For example, for a collinear ferromagnetic spin configuration, in which local spins at sites $i$ and $j$ are given by $\hat{\mathbf{s}}_i = \hat{\mathbf{s}}_j = -\mathrm{sgn}(J_{sd}) \hat{\mathbf{n}}_0$, the effective tunneling reduces to
\begin{equation}
    T^\mathrm{eff}_{ij} = \frac{t_0}{2}
    ( \sigma_0 + \hat{\mathbf{n}}_0\cdot \boldsymbol{\sigma}).
    \label{appendixEq:effective_tunneling.strong_Jsd.parallel_spins}
\end{equation}
The effective tunneling satisfies the condition in Eq.~\eqref{dm_coupling_condition}, which can give rise to a Dzyaloshinskii-Moriya-like Josephson coupling.
In contrast to the cases discussed in Sec.~\ref{sec:josephson_three_sublattice} which consider $|J_{sd}| \ll |t_0|$, strong $s$-$d$ exchange can give rise to comparable amplitudes of the Josephson couplings $J_{nm}$, $\mathbf{D}_{nm}$, and $\Gamma_{nm}$ and can further promote a spatially inhomogeneous $d$ vector texture.
This regime can be pertinent to systems such as proximitized Mn$_3$Ge, in which the $s$-$d$ coupling can be comparable to the hopping amplitude~\cite{Chen2014, Kimata2019}.

As an illustrative example, consider Josephson tunneling in the presence of a ferromagnetic spin texture, in which the tunneling is given by that in Eq.~\eqref{appendixEq:effective_tunneling.strong_Jsd.parallel_spins}.
From Eq.~\eqref{josephson_couplings_decoupled_spin_orbital_expansion},
the Josephson couplings are given by
\begin{subequations}
    \begin{align}
        J_{nm} 
        &\approx
        - \Delta_0 \frac{t_0^2}{2W^2}
        N_{nm},
        \\
        \mathbf{D}_{nm}
        & \approx
        i\Delta_0 \frac{t_0^2}{2W^2}
        N_{nm}
        \hat{\mathbf{n}}_0,
        \\
        \Gamma_{nm}^{ab}
        &\approx
        \Delta_0  \frac{t_0^2}{2 W^2} N_{nm} \hat{n}_0^a \hat{n}_0^b,
    \end{align}
\end{subequations}
in which $N_{nm}$ is the dimensionless length of the interface between superconducting grains $n$ and $m$.
Here, the Heisenberg-like and DM-like Josephson couplings are of similar magnitude.
Following Sec.~\ref{sec:josephson_diode}, the diode efficiency for unitary pairing orders is given by
\begin{equation}
    \eta \approx
    \frac{N_{nm}\Delta_0t_0^4 (\hat{\mathbf{d}}_n \cdot \hat{\mathbf{d}}_m)^2}{W^4 |\tilde{\mathcal{J}}_{nm}|} |\sin (2\phi_1)|,
\end{equation}
in which we have taken $I_2 \approx (e/\hbar) \Delta_0 t_0^4/W^4$ and $\phi_2 \approx 0$ for simplicity, and $\tilde{\mathcal{J}}_{nm}$ is given in  Eq.~\eqref{tilde_J.phase_independent}.
Here, the phase difference in the first order critical current is given by
\begin{equation}
    \tan \phi_1 \approx
    \frac{\hat{\mathbf{n}}_0 \cdot (\hat{\mathbf{d}}_n \times \hat{\mathbf{d}}_m)}{\hat{\mathbf{d}}_n \cdot \hat{\mathbf{d}}_m - (\hat{\mathbf{n}}_0\cdot\hat{\mathbf{d}}_n)
    (\hat{\mathbf{n}}_0\cdot\hat{\mathbf{d}}_m)}.
    \label{appendixEq:diode_efficiency_FM_interface}
\end{equation}
For collinear $d$ vectors, this yields $\phi_1 = 0$ and thus a vanishing diode effect.
However, for noncollinear $d$ vectors, this leads to finite phase offset $\phi_1$, resulting in a finite diode effect, with the diode efficiency largely being determined by the relative alignment of $d$ vectors.
In contrast to the diode efficiency in Eq.~\eqref{diode_efficiency_three_sublattice}, which is primarily determined by the underlying exchange field, the diode efficiency in Eq.~\eqref{appendixEq:diode_efficiency_FM_interface} is primarily dependent on the relative orientation of $d$ vectors at the two sides of the junction.
For example, consider the case that $\hat{\mathbf{d}}_n \cdot \hat{\mathbf{d}}_m = \cos \theta$, and both $\hat{\mathbf{d}}_n$ and $\hat{\mathbf{d}}_m$ are orthogonal to $\hat{\mathbf{n}}_0$ such that $\hat{\mathbf{n}}_0\cdot (\hat{\mathbf{d}}_n \times \hat{\mathbf{d}}_m) = \sin \theta.$
It follows that
\begin{equation}
    \eta_\mathrm{max} \approx
    \frac{2 t_0^2}{W^2} \cos^2 \theta |\sin 2\theta|,
\end{equation}
which has a maximum of $\eta \approx \frac{3 \sqrt{3}}{4} \frac{t_0^2}{W^2}$ at $\theta = \pm \frac{\pi}{6}, \pm \frac{7 \pi}{6}$.

\section{Fermi surface spin texture and pairing correlations in the three-sublattice system}

\label{appendix:spin_texture_and_pairing_correlations}

In this appendix, we further analyze the Fermi surface spin textures and the pairing correlations for the three-sublattice systems on triangular and kagome lattices introduced in Sec.~\ref{sec:models_and_tunneling}.

\subsection{System on kagome lattice}

\begin{figure}
    \centering
    \includegraphics[width= \linewidth]{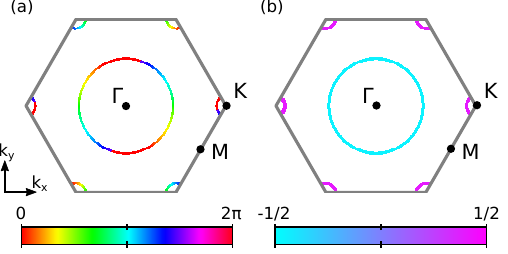}
    \caption{
    Fermi surfaces for itinerant electrons in the $s$-$d$ model on the kagome lattice.
    Plotted is the in-plane azimuthal angle (a) and out-of-plane polarization (b), with the same coloring scheme as Fig.~\ref{fig:FS_spin_texture}.
    Parameters are chosen as $J_{sd} = |t_0|$ and $\mu = 0.2 |t_0|$.
    The three sublattice spins are parameterized by $\varphi_0 = 0,$ $\nu = -1,$ and $\theta_0 = \pi/3,$ corresponding to finite scalar spin chirality in Eq.~\eqref{appendixEq:spin_chirality}.
    }
    \label{AppendixFig:FS_spin_texture.kagome.alt}
\end{figure}

For the $s$-$d$ model  on the kagome lattice, the normal-state band structure  preserves inversion symmetry and features spin-split Fermi surfaces, as discussed in Sec.~\ref{subsec:three-sublattice_models}.
Consider the $120^\circ$-ordered configuration of local $d$ electron spins, as depicted in, for example, Fig.~\ref{fig:tunneling}(a).
The spins of itinerant $s$ electrons exhibit a coplanar chiral texture on the Fermi surface, with opposite chirality for the two spin-split Fermi surfaces.
For surfaces about the $K$ or $M$ points, the winding number is $\pm 1,$ and the Fermi surface about $-K$ or $-M$ respectively has the opposite winding number, as seen in Fig.~\ref{fig:FS_spin_texture}(a).
For Fermi surfaces about the $\Gamma$ point, the winding number is $\pm 2,$ with the spin texture having even partial wave symmetry, as seen in Fig.~\ref{AppendixFig:FS_spin_texture.kagome.alt}(a).
In all cases, the sign of the winding is dependent on the vector chirality ($\pm 120^\circ$-ordering) of the local spin moments, with the spin texture being even under inversion with respect to the $\Gamma$ point.

Introducing an out-of-plane canting $\theta_0 \neq 0$ to the underlying local moments leads to nonvanishing spin chirality,
\begin{equation}
    \chi_{abc} = 
    \mathbf{s}_a \cdot(\mathbf{s}_b \times \mathbf{s}_c) =
    \nu
    \frac{3\sqrt{3}}{16}
    \cos^2\theta_0 \sin \theta_0.
    \label{appendixEq:spin_chirality}
\end{equation}
Consequently, the itinerant electrons gain an out-of-plane spin component.
Shown in Fig.~\ref{AppendixFig:FS_spin_texture.kagome.alt}(b), spin-split Fermi surfaces about the $\Gamma$ and $K$ points have opposite out-of-plane polarization.
A similar picture exists for Fermi surfaces about $K$ or $M$ points, in which the spin-split Fermi surfaces with opposite chirality have opposite out-of-plane polarization for $\theta_0 \neq 0$.
Nonetheless, even in the presence of finite scalar spin chirality, the system on the kagome lattice preserves inversion, with states at $\mathbf{k}$ and $-\mathbf{k}$ having the same local spin texture.

\begin{figure}
    \centering
    \includegraphics[width=\linewidth]{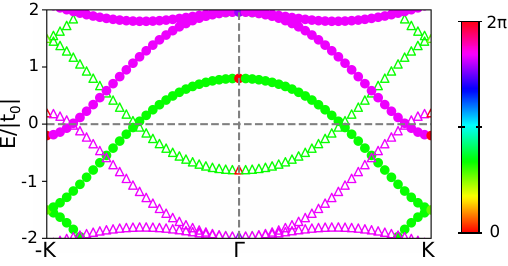}
    \caption{Dispersion along the $\Gamma$-$K$ symmetry line for the $s$-$d$ system on the kagome lattice in the absence of superconductivity ($\Delta_0 = 0$).
    Plotted are the particle-like bands (filled circles) and hole-like bands (open triangles), which are colored on a continuum according to the azimuthal angle of the in-plane spin.
    Parameters are  $\mu = 0.2 |t_0|$ and $J_{sd} = |t_0|$.
    }
    \label{AppendixFig:dispersion_kagome_lattice}
\end{figure}

As discussed in Sec.~\ref{subsec:pairing_correlations.sd_model}, 
proximitized spin singlet superconducting pairing correlations are suppressed  for states at the Fermi surface due to the even-parity spin texture.
Namely, eigenstate $\psi_{\mathbf{s}}(\mathbf{k}, E)$ with momentum $\mathbf{k}$, energy $E(\mathbf{k})$, and spin $\mathbf{s}(\mathbf{k})$ is related by inversion to state $\psi_{\mathbf{s}}(-\mathbf{k}, E)$.
This results in an even-parity spin texture over the Fermi surface, as shown in 
Fig.~\ref{AppendixFig:FS_spin_texture.kagome.alt}.
Consequently, states at the Fermi surface are incompatible with proximitized zero center-of-mass momentum singlet pairing, 
\begin{math}
    |\Psi_\mathrm{sing}\rangle
    =
    ({|\!\uparrow \downarrow\rangle}
    - 
    {|\!\downarrow \uparrow \rangle}
    )/\sqrt{2},
\end{math} and do not open a gap, as seen in Fig.~\ref{fig:pairing_correlations}(a).
Rather, finite pairing correlations for states at the Fermi surface can arise from spin triplet pairing channels.
For zero center-of-mass momentum Cooper pairing, the even-parity symmetry of the Fermi surface spin texture favors spin-polarized superconductivity, with 
\begin{math}
    |\Psi_\mathrm{trip}(\mathbf{k})\rangle
    =
    |\mathbf{s}(\mathbf{k}) \mathbf{s}(\mathbf{k})\rangle.
\end{math}
This leads to a nonunitary $d$ vector in which the spin of the Cooper pair $\hat{\mathbf{S}}_\mathrm{pair}(\mathbf{k}) \propto i\hat{\mathbf{d}}(\mathbf{k}) \times \hat{\mathbf{d}}^*(\mathbf{k})$ is in the local direction of  $\hat{\mathbf{s}}(\mathbf{k})$.
The nonunitary spin-polarized pairing correlations are applicable to systems with or without finite scalar spin chirality and follows from the inversion symmetry of the system on the kagome lattice.

Moreover, in the presence of proximitized spin singlet gap function, zero center-of-mass pairing correlations correspond to interband pairing at higher energies.
In Fig.~\ref{AppendixFig:dispersion_kagome_lattice} we plot the spin texture of eigenstates for the $s$-$d$ system in Eq.~\eqref{hamiltonian_sd} on the kagome lattice along the $\Gamma$-$K$ symmetry line.
States $\psi_{\mathbf{s}}(\mathbf{k}, E)$ and $\psi_{\mathbf{s}'} (-\mathbf{k}, -E)$ for finite $E$ can have different spins, $\hat{\mathbf{s}} \neq \hat{\mathbf{s}}'.$
Although intraband zero-momentum singlet pairing is suppressed by the even-parity spin texture, finite pairing correlations can still arise through interband pairing matrix elements between states whose spin expectation values are not parallel.
As such, higher energy states admit an admixture of spin singlet and spin triplet pairing correlations
from a proximitized spin singlet superconducting gap function.


\subsection{System on triangular lattice}

\begin{figure}[b]
    \centering
    \includegraphics[width= \linewidth]{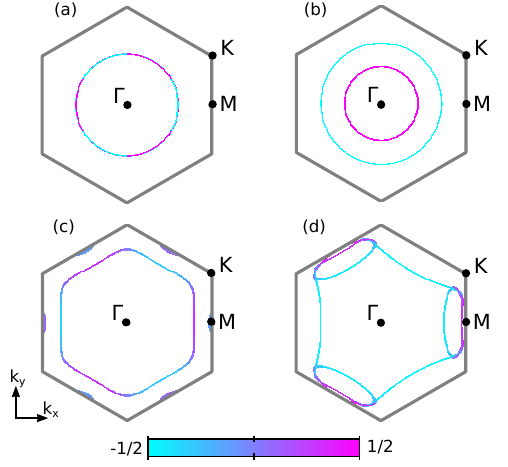}
    \caption{
    Fermi surfaces for itinerant electrons in the $s$-$d$ model on the triangular lattice.
    Doping levels are chosen to realize doubly-degenerate Fermi surfaces about the $\Gamma$ points (a-b), or nondegenerate Fermi surfaces (c-d), with $\mu = -4 |t_0|$ and $\mu = 0$, respectively.
    \textbf{(a,c)}~Spin texture when coupled to coplanar antiferromagnetic spin configuration ($\theta_0 = 0$).
    The two surfaces about the $\Gamma$ point are doubly degenerate with opposite spin polarization, but each displaying odd-parity antiferromagnetic spin texture over the Fermi surface.
    \textbf{(b,d)}~Spin texture when coupled to a noncoplanar spin texture ($\theta_0 = \pi/3$).
    Parameters are chosen as $J_{sd} = |t_0|$, $\lambda_z = 0$, $\varphi_0 = 0$, and $\nu = -1$ for both plots.
    }
    \label{AppendixFig:FS_spin_texture.triangular.alt}
\end{figure}

For itinerant electrons coupled to a $120^\circ$-ordered spin texture on the triangular lattice, the Fermi surfaces display an odd-parity antiferromagnetic texture.
Unlike that of the kagome lattice, the $s$-$d$ model on the triangular lattice is not invariant under inversion.
For Fermi surfaces about the $K$ or $M$ symmetry points, the spin texture is odd under inversion symmetry and is fully polarized in the $z$-direction, as seen in Fig.~\ref{fig:FS_spin_texture}(b).
For Fermi surfaces about the $\Gamma$ point, the spin texture is similarly polarized out-of-plane and has odd partial wave symmetry, as seen in Figs.~\ref{AppendixFig:FS_spin_texture.triangular.alt}(a) and (c).
As described in the main text, this can lead to equal-spin zero center-of-mass momentum pairing.

For finite scalar spin chirality, the spin texture will remain polarized in the $z$-direction but gain a net spin polarization.
For Fermi surfaces which are doubly degenerate about the $\Gamma$ point [Fig.~\ref{AppendixFig:FS_spin_texture.triangular.alt}(a)], the coplanar state exhibits an odd-parity antiferromagnetic spin texture with $f$-wave symmetry when coupled to a coplanar spin texture ($\theta_0 = 0$).
For a noncoplanar spin texture ($\theta_0 \neq 0$), the degeneracy is lifted, resulting in two Fermi surfaces with opposite spin polarization,
as seen in Fig.~\ref{AppendixFig:FS_spin_texture.triangular.alt}(b).
Alternatively, in the case of a nondegenerate Fermi surface about the $\Gamma$ point [Fig.~\ref{AppendixFig:FS_spin_texture.triangular.alt}(c)], 
introducing finite spin chirality will lead to net spin polarization, as seen in Fig.~\ref{AppendixFig:FS_spin_texture.triangular.alt}(d).
In both cases,
zero center-of-mass momentum pairing in presence of finite spin chirality    corresponds to  spin-polarized Cooper pairing, with the spin of the Cooper pair pointing out-of-plane, $i \hat{\mathbf{d}}(\mathbf{k}) \times \hat{\mathbf{d}}^*(\mathbf{k}) = \pm \hat{z}$.

In summary, the systems on the kagome and triangular lattices support qualitatively distinct zero center-of-mass momentum pairing channels due to the symmetry of their Fermi surface spin textures.
Specifically, the kagome lattice realizes an even-parity spin texture that suppresses intraband singlet pairing but favors nonunitary triplet pairing, independent of whether the underlying local moments have finite scalar spin chirality.
The system on the triangular lattice realizes an odd-parity collinear antiferromagnetic texture that permits equal-spin unitary pairing when coupled to coplanar spins, but favors nonunitary pairing when coupled to a noncoplanar spin texture.


%

\end{document}